\documentclass[12pt]{article}
\usepackage{amssymb,amsfonts,latexsym}
\title
{A discrete leading symbol and spectral asymptotics
for natural differential operators}
\author{Ivan Avramidi and Thomas Branson}
\date{August 31, 2000 (revised May 2, 2001)}

\begin{document}

\maketitle

\def\beq{\begin{equation}}
\def\eeq{\end{equation}}
\def\bear{\begin{array}}
\def\eear{\end{array}}

\newcommand{\nnn}[1]{(\ref{#1})}

\newtheorem{theorem}{Theorem}[section]
\newtheorem{lemma}[theorem]{Lemma}
\newtheorem{proposition}[theorem]{Proposition}
\newtheorem{definition}[theorem]{Definition}
\newtheorem{conjecture}[theorem]{Conjecture}
\newtheorem{corollary}[theorem]{Corollary}
\newtheorem{example}[theorem]{Example}
\newtheorem{remark}[theorem]{Remark}

\newcommand{\tfrac}[2]{{\textstyle{\frac{#1}{#2}}}}
\newcommand{\dfrac}[2]{{\displaystyle{\frac{#1}{#2}}}}

\newcommand{\hook}{\raisebox{-0.35ex}{\makebox[0.6em][r]
{\scriptsize $-$}}\hspace{-0.15em}\raisebox{0.25ex}{\makebox[0.4em][l]{\tiny
$|$}}}

\newenvironment{proof}[1]{\begin{trivlist} \item[] {\em #1}. }%
{\hfill $\Box$ \end{trivlist}}

\newcommand{\ul}[1]{\underline{#1}}
\newcommand{\ol}[1]{\overline{#1}}
\newcommand{\text}[1]{{\rm{#1}}}

\def\sideremark#1{\ifvmode\leavevmode\fi\vadjust{\vbox to0pt{\vss
 \hbox to 0pt{\hskip\hsize\hskip1em
 \vbox{\hsize3cm\tiny\raggedright\pretolerance10000
 \noindent #1\hfill}\hss}\vbox to8pt{\vfil}\vss}}}%
                                                   %
\newcommand{\edz}[1]{\sideremark{#1}}

\def\dpsA{{\bf k}}
\def\dpsB{{\bf K}}
\def\dpsC{{\bf J}}
\def\dpsD{{\bf j}}
\def\tens{{\rm tens}}

\def\eye{\sqrt{-1}}
\def\negthickspace{{\!\!}}
\def\sgn{{\rm sgn}}
\def\ta{\tilde\a}
\def\tb{\tilde\b}
\def\tc{\tilde c}

\let\a=\alpha
\let\b=\beta
\let\c=\chi
\let\d=\delta
\let\D=\Delta
\let\e=\varepsilon
\let\f=\varphi
\let\g=\gamma
\let\G=\Gamma
\let\h=\ell
\let\i=\iota
\let\k=\kappa
\let\l=\lambda
\let\L=\Lambda
\let\m=\mu
\let\n=\nu
\let\N=\nabla
\let\p=\pi
\let\P=\Pi
\let\r=\rho
\let\s=\sigma
\let\S=\Sigma
\let\t=\tau
\let\u=\theta
\let\ups=\Upsilon
\let\Ups=\Upsilon
\let\w=\omega
\let\W=\Omega
\let\x=\xi
\let\y=\psi
\let\Y=\Psi
\let\z=\zeta

\let\implies=\Rightarrow

\def\beven{b_{{\rm e}}}
\def\deven{d_{{\rm e}}}
\def\dself{D_{{\rm self}}}
\def\teven{t_{{\rm even}}}
\def\uself{\Upsilon_{{\rm self}}}

\def\bbC{{\mathbb C}}
\def\bbD{{\mathbb D}}
\def\bbN{{\mathbb N}}
\def\bbR{{\mathbb R}}
\def\bbS{{\mathbb S}}
\def\bbT{{\mathbb T}}
\def\bbV{{\mathbb V}}
\def\bbW{{\mathbb W}}
\def\bbZ{{\mathbb Z}}

\def\cA{{\cal A}}
\def\cAe{{\cal A}_{{\rm e}}}
\def\cD{{\cal D}}
\def\cE{{\cal E}}
\def\cF{{\cal F}}
\def\cG{{\cal G}}
\def\cGe{{\cal G}_{{\rm e}}}
\def\cH{{\cal H}}
\def\cP{{\cal P}}
\def\cR{{\cal R}}
\def\cS{{\cal S}}
\def\cT{{\cal T}}
\def\cTe{{\cal T}_{{\rm e}}}
\def\cV{{\cal V}}
\def\ctS{\tilde{\cal S}}
\def\cW{{\cal W}}

\def\gso{{\mathfrak{so}}}

\let\da=\downarrow
\let\lra=\leftrightarrow
\let\ua=\uparrow
\let\os=\oplus
\let\ot=\otimes
\let\bop=\bigoplus

\def\card{{\rm card}}
\def\eig{{\rm eig}}
\def\endo{{\rm End}}
\def\eye{\sqrt{-1}}
\def\fcn{{\rm fcn}}
\def\hom{{\rm Hom}}
\def\id{{\rm id}}
\def\ind{{\rm Ind}}
\def\ort{{\rm O}}
\def\pari{{\rm par}}
\def\pin{{\rm Pin}}
\def\proj{{\rm Proj}}
\def\so{{\rm SO}}
\def\spin{{\rm Spin}}
\def\sym{{\rm sym}}
\def\SYM{{\rm SYM}}
\def\tfs{{\rm tfs}}
\def\TFS{{\rm TFS}}
\def\tf{\tfrac12}

\def\acirc{\stackrel\circ\alpha}
\def\ath{a^{\underline{{\rm th}}}}
\def\ci{C^\infty}
\def\circe{{\scriptstyle{\circ}}}
\def\dc{\hbox{$\nabla\negthickspace\negthickspace /\;$}}
\def\pth{p^{\underline{{\rm th}}}}
\def\spn{\spin(n)}
\def\swsyu{\z_u(\x)^*\z_u(\x)}
\def\swsyun{\z_u(\eta)^*\z_u(\eta)}
\def\swsyT{\z_1(\x)^*\z_1(\x)}
\def\swsyH{\z_2(\x)^*\z_2(\x)}
\def\swsyE{\z_3(\x)^*\z_3(\x)}
\def\swu{G_u^*G_u}

\def\cSe{{\cal A}_{\rm e}}

\begin{center}{\em Dedicated to the memory of Irving Segal}
\end{center}

\begin{abstract} We initiate a systematic study of natural differential
operators in Riemannian geometry whose leading symbols are not of
Laplace type.  In particular, we define a discrete leading symbol
for such operators which may be computed pointwise, or from 
spectral asymptotics.  We indicate how this can be applied to
the computation of another kind of spectral asymptotics, namely
asymptotic expansions of fundamental solutions, and to the 
computation of conformally covariant operators.
\end{abstract}
\maketitle

\section{Introduction} 

In this paper, we would like to set the stage for a better
understanding of natural differential operators in Riemannian (and
Riemannian spin) geometry whose whose leading symbols are not simply
powers of $|\x|^2$.  Such operators or leading symbols have been
called, in various contexts, {\em non-minimal}, {\em nonscalar}, {\em
non-Laplace type}, and even {\em exotic}.  Among the potential
applications are the computation of resolvent and heat operator
asymptotics of elliptic operators with nonscalar leading symbol, and
the computation of explicit formulas for conformally invariant
differential operators.  Our principal tool will be an assignment of a
finite-dimensional commutative algebra $\cA(\l)$ to each irreducible
SO$(n)$ or Spin$(n)$ bundle $\bbV(\l)$.  (The label $\l$ is explained
below.)  This algebra simultaneously encodes information on the
spectrum of the leading symbol (an operator on a
finite-dimensional space), and spectral asymptotics of the realization
of a natural differential operator on the standard sphere $S^n$
(an operator in an infinite-dimensional space).  Thus
it relates global and pointwise information.  A version of this
viewpoint was used in \cite{tbkato} to get sharp improved Kato
constants for solutions of natural first-order elliptic systems on
Riemannian (or Riemannian spin) manifolds.  These constants are
essentially bottom eigenvalues of certain natural symbols.  For other
applications, for example to the computation of spectral invariants
of natural differential operators with nonscalar leading 
symbol, the symbol's complete spectral resolution is required,
and this paper provides that information.

\section{Principal symbols and representation theory}

\subsection{Foundations}\label{Foundations}
Let $H(n)$ be $\so(n)$ or $\spin(n)$, and let $M$ be an $n$-dimensional
$H(n)$ manifold.  That is, if $H(n)=\so(n)$, we require $M$ to be oriented 
and Riemannian; if $H(n)=\spin(n)$, we require $M$ to a a Riemannian spin
manifold.  
Let $\bbV(\l)=\cF\times_\l V$ be the vector bundle
canonically associated to a finite-dimensional irreducible representation
of $H(n)$ and the bundle $\cF$ of $H(n)$-frames (i.e.\ oriented orthonormal
frames, or Riemannian spin frames).  Note that if we have spin structure,
we may take $H(n)$ to be $\spin(n)$ even for $\so(n)$ bundles, since we
may always compose with the covering homomorphism $\spin(n)\to\so(n)$.

Let $(\t,T)$ be the defining representation of $\so(n)$; then
$\bbV(\t)$ is the cotangent (or tangent) bundle.  It is well known
that irreducible $\so(n)$ bundles are $H(n)$-isomorphic to {\em tensor
bundles}; i.e. direct summands of tensor powers of $\bbV(\t)$; in
fact, this is guaranteed by the faithfulness of the representation
$\t$.  Similarly, because the spin representation $(\s,\S)$ (which
splits into two irreducible direct summands $(\s_\pm,\S_\pm)$ in even
dimensions) is faithful, all irreducible $\spin(n)$ bundles are
summands in some tensor power of $\bbV(\s)$.  Since $\s\otimes\s$ and
$\s_\pm\otimes\s_{\pm'}$ are $\spin(n)$ isomorphic to tensor bundles
(in fact, differential form bundles), each {\em proper} $\spin(n)$ bundle
(i.e., each $\spin(n)$ bundle which is not an $\so(n)$ bundle) is
realizable as a direct summand of some
$\s\otimes\t\otimes\cdots\otimes\t$.  That is, each may be realized as
a bundle of tensors with spinor coefficients.
A given $\bbV(\l)$ may not have a distinguished 
real form, so we generally think of our sections as being complex.
It is possible, however, to speak of real cotangent vectors, and this
is important in analytic considerations like those of \cite{sw,tbkato}.

The classical {\em branching rule} gives the direct sum
decomposition an irreducible $H(n)$-module 
$(\l,V)$ when restricted to a copy of $H(n-1)$ which is imbedded
in the standard way.  (On the Lie algebra level, relative to some
orthonormal basis of the defining module, the subalgebra should be
that of matrices living in the upper left $(n-1)\times(n-1)$ block.
To see that nonstandard imbeddings are possible, see \cite{spg}, 
Sec.\ 3.c.)
It is known (see Sec.\ \ref{branchsubsection} 
below) that the branching rule is
{\em multiplicity free}:
\begin{equation}\label{branch}
\l|_{H(n-1)}\cong_{H(n-1)}
\b_1\oplus\cdots\oplus\b_{b(\l)},
\end{equation}
where the $\b_i$ are irreducible representations of $H(n-1)$,
and $\b_i\cong_{H(n-1)}\b_j\iff i=j$.  
We put 
$$
B(\l)=\{\b_1,\ldots,\b_{b(\l)}\}.
$$
In particular, the above defines
a numerical invariant $b(\l)$, the cardinality of the set of 
branches $B(\l)$.

By Weyl's invariant theory and the above remarks on tensor and
tensor-spinor realizations, an $H(n)$-equivariant differential
operator on (sections of) $\bbV(\l)$ is 
built polynomially from the covariant derivative (with respect
to the Levi-Civita or Levi-Civita spin connection $\N$),
the Riemann curvature $R$, the metric $g$ and its inverse $g^\sharp$, 
the volume form $E$, and, if applicable, the
fundamental tensor-spinor $\g$.  Of course, such
objects are not really operators, but functors which assign operators
to $H(n)$ manifolds.  (Similarly, the $\bbV(\l)$ are not bundles, but
functors assigning bundles.)
To be of {\em universal} order 
$p$, i.e.\ order $p$ on every $H(n)$-manifold,
such an operator must have the form
\beq\label{PrincAndLower}
D=D_{\rm princ}+D_{\rm lower},
\eeq
where $D_{\rm princ}$ is a sum of monomials in all the above ingredients
{\em except} $R$, each monomial containing $p$ occurrences of $\N$,
and ${\rm ord}(D_{\rm lower})\le p-1$ on each $H(n)$-manifold.
$R$ must be is missing from the list of ingredients in $D_{\rm princ}$,
to avoid the vanishing of $D_{\rm princ}$ on flat manifolds
(i.e., to avoid contradicting the universality of the order $p$).
The decomposition \nnn{PrincAndLower} is not unique, as commutation
of covariant derivatives produces lower order terms.  Nevertheless,
we may read off from this decomposition the fact that the 
leading symbol $\s_p(D)$, when viewed as a section of 
$\hom(\SYM^p\otimes\bbV(\l),\bbV(\l))$, is {\em parallel}; i.e., annihilated
by the covariant derivative, since $g$, $E$, and $\g$ are.
Here $\SYM^p$ is the bundle of symmetric $p$-tensors.
$\s_p(D)$ is also 
$H(n)$-invariant; that is, it is actually in
$\hom_{H(n)}(\SYM^p\otimes\bbV(\l),\bbV(\l))$.
It is {\em universal}, in the sense of being given by a consistent expression
in $g$, $g^\sharp$, $E$, and, if applicable, $\g$, at all points of
all $H(n)$-manifolds.  
In fact, we get such leading symbols by ``promoting'' to the
bundle setting {\em actions} of the $H(n)$ module $\sym^p$ of
symmetric $p$-tensors
on $(\l,V)$; that is, elements of $\hom_{H(n)}(\sym^p\otimes\l,\l)$.

It will also be useful to speak of the {\em reduced order} of an
$H(n)$-operator or leading symbol.  Given an $H(n)$-symbol $\u(\x)$ of 
homogeneous degree $p$,
write $\u(\x)=|\x|^{2k}\u_{{\rm red}}(\x)$ for $k$ as large as possible;
the {\em reduced symbol} is $\u_{{\rm red}}(\x)$.
The reduced homogeneous degree, 
$p-2k$, is well-defined, and may also be detected
as the degree of the restriction of $\u$ to the unit $\x$-sphere
(i.e., the unit sphere in the module $(\t,T)$);
we shall denote this restriction by $\tilde\u$.
In fact, under the identification of symmetric tensors and 
homogeneous polynomials,
the decomposition into {\em trace-free} tensors of various degrees
corresponds to the decomposition into terms of the form
$$
|\x|^{2k}\cdot{\rm(spherical\ harmonic)}.
$$
If $\u$ has homogeneous order $p$, then
\begin{equation}\label{gradetfs}
\tilde\u=\tilde\u_p+\tilde\u_{p-2}+\cdots+\tilde\u_{0\;{\rm or}\;1},
\end{equation}
where $\tilde\u_q$ corresponds to an element of
$\hom_{H(n)}(\tfs^q\otimes\l,\l)$.

We define the algebra $\cA(\l)$ of $H(n)$-equivariant 
{\em principal symbols} as that generated
by the $\tilde\u$ above.  
The difference between a {\em principal} and a {\em leading} symbol
is that we allow ourselves to add symbols of different orders within
the principal symbol algebra.  
Given a choice of a {\em real} $\x$ on the unit sphere in $(\t,T)$, 
we get a decomposition \nnn{branch}
of the module $\l$ under the $H(n-1)$ subgroup
fixing $\x$.  By Schur's Lemma and the multiplicity free nature of
\nnn{branch}, if 
$\k\in\cA(\l)$, then $\k(\x)$ acts as multiplication by some scalar $\m_i$ 
on each $\b_i$ summand.  
By the invariance of $\k$ and the transitivity of $H(n)$ on $(\t,T)$, 
the eigenvalue list $\m_i$ is independent of the choice of $\x$,
and $\m_i$ is always attached to the $\b_i$ summand.
Thus $\cA(\l)$ is isomorphic to the 
algebra of complex-valued
functions on the finite set $B(\l)$,
and, in particular, is commutative.

\begin{definition} The map $\b_i\mapsto\m_i$ defined above is the
{\em discrete leading symbol}.  That is, we define $\dpsA:\cA(\l)\to{\rm maps}
(B(\l),\bbC)$ by
$$
\dpsA(\k)(\b_i)=\m_i.
$$
If $D$ is an $H(n)$-invariant differential operator of order
$p$ on $\bbV(\l)$, we put
\beq\label{factorA}
\dpsB(D)=\dpsA(\tilde\s_p(D)).
\eeq
\end{definition}

In other words, the discrete leading symbol is the spectral resolution
of a principal symbol, or of the leading symbol, at a given unit $\x$.
The eigenvalues, with multiplicities, are independent of $\x$, and the
eigenspaces move in a predictable way, according to the action of
$H(n)$.

There is a grading of $\cA(\l)$ by order, in which 
an action of $\tfs^p$ falls
into $\cA^p(\l)$.  In particular, we shall give the name $d(\l)$
to the maximal
$p$ for which $\tfs^p$ acts:
$$
d(\l):=\max\{p\mid\hom_{H(n)}(\tfs^p\otimes\l,\l)\ne 0\}.
$$
The behavior of this grading under multiplication is somewhat involved;
in fact, in view of \nnn{gradetfs}, this is exactly the problem of 
decomposing products of spherical harmonics into sums of 
spherical harmonics.
\nnn{gradetfs} does, however, allow a simple description of the 
behavior of multiplication under a coarser grading.  Let
$\cA_0(\l)$ (resp.\ $\cA_1(\l)$) be the direct sum of the $\cA^p(\l)$
over even (resp.\ odd) $p$.  Then
$$
\cA_i(\l)\cA_j(\l)\subset\cA_{i+j}(\l),
$$
where the addition in the subscripts is modulo 2.  It is often the
case that the odd part vanishes (see Theorem \ref{gradspan} below).  
Since the 
reduction of any $|\x|^{2k}$ is 1, any principal symbol which
is purely even or odd,
i.e.\ any symbol $\k$ in $\cA_0(\l)$
or in $\cA_1(\l)$, will be {\em represented by} an actual differential
operator; that is, there will be some invariant $D$ of some order $p$
with $\tilde\s_p(D)=\k$.  Indeed, to get such a $D$, first get an
invariant {\em homogeneous} polynomial $\u$ in $\x$ with $\tilde\u=\k$;
then replace each $\x$ by an 
$-\eye\N$ in the tensorial expression for $\u$.

We summarize some of the above considerations in:

\begin{proposition}\label{foundations}
The algebra $\cA(\l)$ of principal symbols on $\bbV(\l)$
is isomorphic to the 
algebra of complex-valued functions on the finite set $B(\l)$.
In particular, it is 
commutative, and is generated by $b(\l)$ fundamental projections.
Each principal symbol in $\cA_0(\l)$ or in $\cA_1(\l)$ is the reduced
leading
symbol of an $H(n)$-invariant operator 
of order at most $d(\l)$.
\end{proposition}

To see the fundamental projections in a very familiar (but deceptively
simple) case, consider
the bundle $\L^k$ of differential $k$-forms, $0<k<(n-2)/2$.  
The fundamental projections are $\i(\x)\e(\x)$ and $\e(\x)\i(\x)$,
where $\e$ and $\i$ are exterior and interior multiplication.  Here
\begin{itemize}
\setlength{\itemsep}{-1ex}
\item there are 2 fundamental projections;
\item each has degree 2 in $\x$;
\item each is represented by a differential operator (the Hodge operators
$\d d$ and $d\d$);
\item the orthogonality of the projections persists on the operator level:
$\d dd\d=d\d\d d=0$.
\end{itemize}
By way of contrast, in the general $\bbV(\l)$ case,
\begin{itemize}
\setlength{\itemsep}{-1ex}
\item there are $b(\l)$ fundamental projections, and $b(\l)$ can be arbitrarily
large, depending on $\l$;
\item the maximal degree of a projection in $\x$ is $d(\l)$, which can be 
arbitrarily large, depending on $\l$;
\item the fundamental projections need not be represented by differential 
operators;
\item even if they are, there is generally no choice of such operators
$D_i$ with $D_iD_j=0$ for $i\ne j$.
\end{itemize}

\subsection{A calculation involving translation and tensor products of
representations}
A standard computation from the theory of induced representations actually
leads immediately to the structure of the principal symbol algebra.
Let $1$ be the trivial
representation of $H(n-1)$, and let $(\l,V)$ and $(\m,W)$ be finite-dimensional
representations of $H(n)$.  
We claim that
\begin{equation}\label{ind1}
\hom_{H(n)}((\ind_{H(n-1)}^{H(n)}1)\otimes\l,\m)\cong
\hom_{H(n-1)}(\l|_{H(n-1)},\m|_{H(n-1)}).
\end{equation}
Indeed,
$$
(\ind_{H(n-1)}^{H(n)} 1)\otimes 
\l\cong_{H(n)}\ind_{H(n-1)}^{H(n)}(1\otimes \l|_{H(n-1)})
\cong_{H(n)}\ind_{H(n-1)}^{H(n)}(\l|_{H(n-1)});
$$
this is a form of the {\it translation principle}.
By Frobenius Reciprocity, 
$$
\hom_{H(n)}(\ind_{H(n-1)}^{H(n)}(\l|_{H(n-1)}),\m)
\cong\hom_{H(n-1)}(\l|_{H(n-1)}\,,\m|_{H(n-1)}),
$$
proving \nnn{ind1}.
In particular, if $\l=\m$, we get
$$
\hom_{H(n)}((\ind_{H(n-1)}^{H(n)}1)\otimes \l,\l)
\cong\endo_{H(n-1)}(\l|_{H(n-1)}).
$$
(This calculation can actually be done with any reductive pair
in place of $(H(n),H(n-1))$.)

The significance of this in the present context is as follows.
The expansion of 
$\ind_{H(n-1)}^{H(n)}1$ is exactly the expansion of $\so(n)$-finite 
functions on the sphere $S^{n-1}$
into spherical harmonics of
different degrees $p$:
\begin{equation}\label{IndAndTFS}
\ind_{H(n-1)}^{H(n)}1\cong_{\so(n)}\bop_{p=0}^\infty
\TFS^p.
\end{equation}
This shows that the leading symbol algebra must isomorphic to 
$\endo_{H(n-1)}(\l)$, which, in view of the multiplicity-free nature 
of the branching rule, is another realization of the complex-valued
functions on $B(\l)$.

\subsection{The selection rule and Stein-Weiss operators} 
An important class of $H(n)$-equivariant differential operators on $\bbV(\l)$
are the {\em generalized gradients} of
Stein and Weiss \cite{sw}.
The starting point is the {\em
selection rule}, 
which describes the $H(n)$-decomposition of $\t\otimes\l$.  (Recall that
$\t$ is the defining representation of $\gso(n)$.)
As it happens, this decomposition, like the branching rule, is 
also multiplicity free:
$$
\t\ot\l\cong_{H(n)}\m_1\os\cdots\os\m_{N(\l)}\,,
$$
where the $\m_u$ are irreducible representations of $H(n)$, and 
$$
\m_u\cong_{H(n)}\m_v\iff u=v.
$$
In particular, this defines a numerical invariant $N(\cdot)$
on $\hat H(n)$.
On the bundle level,
\begin{eqnarray*}
T^*M\ot\bbV(\l)&\cong_{H(n)}&\bbV(\t)\ot\bbV(\l) \\
&\cong_{H(n)}&\bbV(\s_1)\os\ldots\os\bbV(\s_{N(\l)}).
\end{eqnarray*}
The covariant derivative $\N$ carries sections of $\bbV(\l)$ to 
sections of $T^*M\otimes\bbV(\l)$.
Because the selection rule is multiplicity free, 
we may project $\N$ onto the unique summand
of covariance type $\m_u$ to obtain our generalized gradient:
$$
G_{\l\s_u}=G_u=\proj_u\circe\N.
$$
Up to normalization and up to $H(n)$-isomorphic realization of
bundles, some examples of gradients or direct sums of gradients are
the exterior derivative on forms, the conformal Killing operator on
vector fields, the Dirac operator, the twistor operator, and the
Rarita-Schwinger operator.  In fact, every first-order
$\spin(n)$-equivariant differential operator is a direct sum of
gradients \cite{feg}.  In particular, the formal adjoint of a
gradient is also a gradient.  This immediately gives us access to a
very important class of operators carrying $\bbV(\l)$ to itself, namely the
$G_u^*G_u$.  For some $\l$ in odd dimensions, there is a {\em
self-gradient}; that is, some $\m_u$ is $H(n)$-isomorphic to $\l$
itself, so that there is a natural first-order operator
$D_{{\rm self}}$ carrying sections of $\bbV(\l)$
to sections of $\bbV(\l)$.  (The most familiar examples are the Dirac
operator on spinors, the Rarita-Schwinger operator on twistors,
and the operator $\star d$ on $(n-1)/2$-forms.)
We shall show below that the leading symbols of the $G_u^*G_u$ and,
when it exists, $D_{{\rm self}}$, generate the principal symbol
algebra $\cA(\l)$.  

\begin{remark}\label{SelfG}
{\rm In case $\l$ admits a self-gradient, 
that is if $\l$ itself is a selection rule target $\m_u$ for $\l$,
the target of $G_u$ is ``born'' as
a subbundle $\bbW$ of $T^*M\otimes\bbV(\l)$
with $\bbW\cong_{\spin(n)}\bbV(\l)$.  If we would like to use the same
realization of $\bbV(\l)$ as both source and target bundle for
a realization $D_{{\rm self}}$ of $G_u$, 
we need a choice of normalization.
First, normalize the Hermitian inner product on $T^*M\otimes\bbV$
so that $|\xi\otimes v|^2=|\xi|^2|v|^2$; then define 
$D_{{\rm self}}$ so that
\beq\label{dselfsq}
D^2_{{\rm self}}=G_u^*G_u.
\eeq
Note that this determines $D_{{\rm self}}$ only up to multiplication
by $\pm 1$.  This sign ambiguity is already apparent in 
the definition of the Dirac operator on spinors as $\g^a\N_a$,
since changing $\g$ to $-\g$ does not disturb the Clifford
relations, nor the relation $\N\g=0$.}
\end{remark}

As shown in \cite{tbsw}, all second-order operators are linear 
combinations of the $G_u^*G_u$, modulo lower-order operators.
The leading symbols of the $G_u^*G_u$, however, are generally not
linearly independent.  In fact, as shown in 
\cite{tbsw}, Theorem 5.10, their leading symbols form a space
of dimension
\begin{equation}\label{t(lambda)}
t(\l):=[(N(\l)+1)/2].
\end{equation}
The full set of linear relations among the leading symbols of
the $G_u^*G_u$ is also given explicitly in \cite{tbsw}, Theorem 5.10.

\subsection{Numerical invariants of bundles}
To recap, we have defined the following numerical invariants of an
irreducible $H(n)$-bundle $\bbV(\l)$.
All of these are really invariants of the underlying representation
$\l$.
\begin{eqnarray*}
\text{number\ of\ selection\ rule\ summands}:&\qquad&N(\l) \\
\text{maximal\ degree\ in\ }\cA(\l):         &\qquad&d(\l) \\
\text{dimension,\ thus\ number\ of\ fundamental\ projections,\ 
of\ }\cA(\l):                                &\qquad&b(\l)  \\
\text{number\ of\ linearly\ independent\ }\s_2(G_u^*G_u):
                                             &\qquad&t(\l).
\end{eqnarray*}
Of these, $t(\l)$ and $N(\l)$ are related by \nnn{t(lambda)}.
We might add the following, to take account of the even/odd grading:
\begin{eqnarray*}
\text{maximal\ degree\ in\ }\cA_i(\l)\ (i=0,1):    &\qquad&d_i(\l) \\
\text{dimension,\ thus\ number\ of\ fundamental\ projections,\ 
of\ }\cA_0(\l):                                &\qquad&b_0(\l)  \\
\dim\cA_1(\l): & \qquad & b_1(\l).
\end{eqnarray*}
So far, we have done nothing explicit to actually
compute these invariants, 
the fundamental projections in $\cA(\l)$, or 
any discrete leading symbols.  We shall now remedy this situation.

\section{The weight game}

\subsection{Parameterization by dominant weights}
Irreducible representations of $\spin(n)$,
and thus irreducible associated $\spin(n)$-bundles, are parameterized
by {\it dominant weights} 
$(\l_1\,,\ldots\,,\l_\h)\in\bbZ^\h\cup(\frac12+\bbZ)^\h$, 
$\h=[n/2]$, satisfying the inequality constraint
\begin{equation}\label{dominant}
\begin{array}{ll}
\l_1\ge\ldots\ge\l_\h\ge 0,\qquad & n\text{\ odd}, \\
\l_1\ge\ldots\ge\l_{\h-1}\ge|\l_\h|\,,\qquad & n\text{\ even}.
\end{array}
\end{equation}
The dominant weight $\l$ is the highest weight of the corresponding
representation.  
The representations which factor through $\so(n)$ are exactly those with
$\l\in\bbZ^\h$. 

Note that in the previous sections, we have used
$\l$ as a notation for 
the representation itself.  
This is a standard abuse of notation, which we shall continue:
the highest weight parameter
of an irreducible representation will be used as a synonym
for the representation.
We shall also denote by $\c(n)$ the set of dominant $H(n)$ weights.

\subsection{The selection rule} 
We shall discuss several familiar examples of bundles and identify
their highest weights below.  One important highest weight is that of
the defining representation $\t$, namely $(1,0,\ldots,0)$.  With this,
we can explicitly 
describe the {\it selection rule} mentioned above.  If $\l$ is
an arbitrary irreducible representation of $\spin(n)$, then
$$
\t\ot\l\cong_{H(n)}\s_1\os\cdots\os\s_{N(\l)}\,,
$$
where the $\s_u$ are distinct,
and a given $\s$ appears 
if and only if
$\s$ is a dominant weight and 
\begin{eqnarray}
&\s=\l\pm e_a\,,\text{\ some\ }a\in\{1,\ldots,\h\},\qquad 
\underline{\text{or}}\label{selection} \\
&n\text{\ is\ odd,\ }\l_\h\ne 0,\ \s=\l.\label{selectionExc}
\end{eqnarray}
Here $e_a$ is the $\ath$ standard basis vector in $\bbR^\h$.
The selection rule follows immediately from the Brauer-Kostant
formula; see \cite{brauer,kost}.
We shall use the notation
$$
\l\lra\s
$$
for the selection rule: $\l\lra\s$ if and only if $\s$ appears 
as a summand in $\t\ot\l$.
The notation $\lra$ is justified because the relation is symmetric.
In fact, one can see {\it a priori} that $\lra$ must be symmetric:
$\t$ is a real representation, 
and thus is self-contragredient.

\subsection{The branching rule}\label{branchsubsection} 
The explicit {\it branching rule} for the restriction from $H(n)$ to
$H(n-1)$ is as follows.  By changing $n$ to $n-1$ above we have
a parameterization of the irreducible representations of
$H(n-1)$.  The branching rule says that 
for a dominant $H(n)$-weight $\b$, 
$\dim\hom_{H(n-1)}(\b,\l|_{H(n-1)})=0$ or $1$, with
$\dim\hom_{H(n-1)}(\b,\l|_{H(n-1)})=1$ if and only if
\begin{equation}\label{BRANCH}
\l_1-\b_1\in\bbZ\quad\text{and}\quad
\left\{\begin{array}{ll}
\l_1\ge\b_1\ge\l_2\ge\cdots\ge\b_{\h-1}\ge|\l_\h|,\quad & n\text{\ even}, \\
\l_1\ge\b_1\ge\l_2\ge\cdots\ge\b_{\h-1}\ge\l_\h
\ge|\b_\h|,\quad & n\text{\ odd}.
\end{array}\right.
\end{equation}
We use $\l\da\b$ or $\b\ua\l$ as an abbreviation for \nnn{BRANCH}.  We
shall actually also have use for the version of the branching rule that
restricts from $H(n+1)$ to $H(n)$, so notations like
$\l({}_n\!\da_{n-1})\b$ and $\a({}_{n+1}\!\da_n)\l$ will sometimes be
helpful.
In this connection the following observation will be useful.  

\begin{lemma}\label{parents}
For
$\a\in\c(n+1)$, let $\acirc$ be the $[(n-1)/2]$-tuple
$(\a_2\,\ldots,\a_{[(n+1)/2]})$; that is, $\a$ without its first
entry.  Then $\acirc\in\c(n-1)$,
and given $\l\in\c(n)$, 
$\{\acirc\mid\a({}_{n+1}\!\da_n)\l\}=
\{\n\mid\l({}_n\!\da_{n-1})\n\}$.
\end{lemma}

The proof comes directly upon examination of the branching rule.  To
paraphrase, the branching ``offspring'' of $\l\in\c(n)$ are the 
$\acirc$ for the branching ``parents'' $\a$ of $\l$.
The branching rule also allows us to give a formula for the dimension
$b(\l)$ of $\cA(\l)$, based on its identification with $\endo_{H(n-1)}
(\l|_{H(n-1)})$:

\begin{lemma}\label{BranchNumber}
\begin{displaymath}
b(\l)=\dim\cA(\l)
=\left\{\begin{array}{ll}
(\l_{\h-1}-|\l_\h|+1)\prod_{a=1}^{(n-4)/2}(\l_a-\l_{a+1}+1),\qquad
& n\ge 4\text{\ even,} \\
([\l_\h]+1)\prod_{a=1}^{(n-3)/2}(\l_a-\l_{a+1}+1),\qquad
& n\ge 3\text{\ odd.}
\end{array}\right.
\end{displaymath}
\end{lemma}

To paraphrase, $B(\l)$ consists of lattice points in a rectangular
box, 
whose various widths are determined by the spacing between successive
entries of $\l$.  Thus ``steeper'' $\l$ tend to produce larger 
$b(\l)$.  Familiar bundles, like spinors, differential forms, and
trace-free symmetric tensors, tend to have relatively ``flat''
$\l$.

We know that $\cA^p(\l)=0$ for
large $p$.  A more quantitative form of this statement is:

\begin{lemma}\label{Can'tAct}  If $n\ge 3$, then 
$p>2\l_1\Rightarrow\cA^p(V)=0$.  As a result, $d(\l)\le 2\l_1$.
\end{lemma}

\begin{proof}{Proof} 
First note that $\tfs^p\cong_{H(n)}(p,0,\ldots,0)$ (see, e.g., \cite{sw}).
If $\hom_{H(n)}(\tfs^p\ot\l,\l)$ is to be nonzero, we must be able
to realize $\l$ as $(p,0,\ldots,0)+\m$, where $\m$ is a weight
of $V(\l)$.  
(This follows, for example, from the Brauer-Kostant 
formula \cite{brauer,kost}, which
expresses the highest weights in summands of the tensor product
in terms of the highest weight of one factor, together with all weights
of the other factor.)
All components of such a $\m$ must be $\le\l_1$ in absolute
value, since
otherwise, we would have $w\cdot\m$ dominant and 
$(w\cdot\m)_1>\l_1$ for some element $w$ of
the Weyl group.
($w\cdot\m$ is a weight of $V(\l)$ since the Weyl group permutes
the weights of any finite-dimensional representation.)  This gives
$$
\m_1=\l_1-p,\qquad|\m_1|\le\l_1\,,
$$
whence $p\le 2\l_1\,$.
The bound on $d(\l)$ is just a restatement, since $d(\l)$
is the maximal $p$ that occurs.
\end{proof}

\begin{remark}\label{Act2} {\rm 
The basic principle of the proof of Lemma \ref{Can'tAct} may be used
to get more refined information in special situations, without going
through the full tensor product calculation.  For example, if $n$ is
even and $\l=(p,\ldots,p)$ with $p>0$, then $d(\l)<2p$, since
$$
(p,\ldots,p)-(2p,0,\ldots,0)=(-p,p,\ldots,p)
$$
is in the Weyl orbit of $(p,\ldots,p,-p)$, a dominant weight not appearing
in $V(p,\ldots,p)$.}
\end{remark}

The leading symbol of any gradient $G_u$ is a relatively 
familiar object
from representation theory, namely 
the projection of a tensor product onto an
irreducible summand.  
Let $\x$ be a vector from the defining representation, and consider
\begin{eqnarray*}
\tens(\x):\l&\to& \t\ot\l, \\
v&\mapsto&\x\ot v.
\end{eqnarray*}
Compose with the projection onto the $\m_u$ summand of 
$\t\ot\l$ to get a map $\z_u(\xi)$, and use the
functoriality of the associated bundle construction to promote
this to a bundle map
$$
\bbV(\l)\to\bbV(\s_u),
$$
also denoted (in a slight abuse of notation) by $\z_u(\x)$.
Then 
$$
\s_1(G_u)(\x)=\eye\z_u(\x),\qquad\s_1(G_u^*)=-\eye\z_u(\x)^*.
$$
$\z_u$ may also be described by the formula $[G_u,m_f]=
\z_u(df)$, where $m_f$ is multiplication by the $\ci$ function $f$.

Let $\eye\Upsilon_{\text{self}}(\x)$ 
be the leading symbol of the self-gradient,
when it exists.
The following will be a consequence of Theorem \ref{gradspan} below:

\begin{theorem}\label{gen}
$\cA(\l)$ is generated by the restrictions to the unit-$\x$ sphere
(i.e., to the unit sphere bundle in the cotangent bundle) of the
$\z_u(\xi)^*\z_u(\xi)$ unless $\l\lra\l$, in which case
the $\z_u(\xi)^*\z_u(\xi)$ for $\m_u\ne\l$ together with
$\Upsilon_{{\rm self}}(\x)$ generate.  $\cAe(\l)$ is generated,
in all cases, by the $\z_u(\xi)^*\z_u(\xi)$.
\end{theorem}

Note that $\z_{\l\l}(\x)^*\z_{\l\l}(\x)$ coincides, when it exists,
with $\Upsilon_{{\rm self}}(\x)^2$, by \nnn{dselfsq}.

\section{Spectral asymptotics on the sphere}

Let $\a$ be an irreducible 
$H(n+1)$-module (identified with its highest weight label when convenient).
The $H(n+1)$-finite section space $\G(\bbV(\l))$ 
of $\bbV(\l)$ forms the space of the
induced representation in the middle term of the following, and
Frobenius Reciprocity supplies the second $\cong$:
$$
\hom_{H(n+1)}(\Gamma(\bbV(\l)),\a)\cong
\hom_{H(n+1)}(\ind_{H(n)}^{H(n+1)}\l,\a)\cong
\hom_{H(n)}(\l,\a|_{H(n)}).
$$
Thus the $H(n+1)$-finite section space of $\bbV(\l)$ is
\begin{equation}\label{sections}
\Gamma(\bbV(\l))\cong_{H(n+1)}\bop_{\c(n+1)\owns\a\da\l}\a.
\end{equation}
Let $\cV(\a;\l)$ be the subspace of $\Gamma(\bbV(\l))$ which is isomorphic
to $\cV(\a)$; then
\begin{equation}\label{secs}
\Gamma(\bbV(\l))=\bop_{\c(n+1)\owns\a\da\l}\cV(\a;\l).
\end{equation}

For $\l\in\c(n)$, let $\tilde\l$ be the rho-shift of $\lambda$:
$$
\tilde\l=\l+\r_n\,,\qquad 2\r_n=(n-2,n-4,\ldots,n-2\ell).
$$
Notice that for the map 
$$
\c(n+1)\to\c(n-1),\qquad\a\mapsto\acirc,
$$
we have
$$
(\acirc)\tilde{\;}=(\tilde\a)^\circ,
$$
since
$$
\r_{n-1}={\stackrel{\circ}{\r}}_{n+1}\,.
$$

We would like to use the action of differential operators
on these section spaces to 
define {\em another} discrete leading symbol, and 
establish its connection to the one already treated in Sec.\ 
\ref{Foundations}.

\begin{theorem}\label{asymp} Let $D$ be an $H(n)$-invariant operator
on $\bbV(\l)$ of the form
$$
D=D_{{\rm princ}}+D_{{\rm lower}},
$$
where $D_{{\rm princ}}$ is a $p$-homogeneous polynomial in the
$G_u^*G_u$ and, if applicable, $D_{{\rm self}}$, and 
${\rm ord}(D_{{\rm lower}})<p$.  (The homogeneity degree
counts 2 for each $G_u^*G_u$ and 1 for each $D_{{\rm self}}$.)
Then the realization $D_0$ of $D$ on the standard sphere $S^n$
has spectral asymptotics of the form
$$
\eig(D,\a)=\dpsC_p(D)(\stackrel{\circ}{\a})j^p+O(j^{p-1}),
$$
where $\a=(\l_1+j,\acirc)$.
\end{theorem}

\begin{proof}{Proof} The fact that $D$ has an eigenvalue on the
$\a$ summand $\cV(\a;\l)$ of $\G(\bbV(\l))$
is guaranteed by the multiplicity free
nature of the branching rule; i.e. by the fact that there is only
one summand of covariance type $\a$.  The asymptotics for $D_{{\rm princ}}$
are guaranteed by \cite{tbsw}, Theorem 4.1 and \cite{ssgs}, 
Corollary 8.2.
The standard elliptic estimate shows that the addition of 
$D_{{\rm lower}}$ does not disturb things: indeed,
\beq\label{ellest}
|(D_{{\rm lower}}\f,\f)_{L^2}|\le{\rm const}
((\N^*\N)^{p/2}\f,\f)_{L^2}^{(p-1)/p}
\eeq
and $\eig(\N^*\N,\a)=j^2+O(j)$ by \cite{tbhar}, Theorem 1.1.
\end{proof}

Let $\cG^p(\l)$ be the algebra generated by all
$H(n)$-equivariant differential operators of
the type described in 
Theorem \ref{asymp}.

\begin{theorem}\label{gradspan} 
For sufficiently large $k$, the linear transformation 
$\dpsC_{2k}\oplus\dpsC_{2k+1}:\cG^{2k}(\l)\oplus\cG^{2k+1}(\l)
\to{\rm maps}(B(\l),\bbC)$
is onto.  If $n$ is even or $\l_\ell=0$, the linear transformation
$\dpsC_{2k}:\cG^{2k}(\l)
\to{\rm maps}(B(\l),\bbC)$ is onto.
\end{theorem}

\begin{proof}{Proof} 
Note that the definitions immediately give
$$
D\in\cG^p(\l),\;E\in\cG^q(\l)\;\Rightarrow\;\dpsC_{p+q}(DE)
=\dpsC_p(D)\dpsC_q(E),
$$
and if $L$ is $H(n)$-invariant of order $<2k$, 
$$
\dpsC_{2k}((\N^*\N)^k+L)=1,
$$
where ``1'' is the function on $B(\l)$ which is identically 1.

By \cite{tbsw}, Theorem 4.1, there are nonzero constants
$c_u=c_{\l\m_u}$ and 
$\tilde c_u=\tilde c_{\l\m_u}$ such that
\beq\label{TheSpectrum}
\eig(G_u^*G_u,\a)=c_u\prod_{a=1}^{[(n+1)/2]}(\tilde\a_a^2-s_u^2)=
\tilde c_u\prod_{a\in\cT(\l)}(\tilde\a_a^2-s_u^2),
\eeq
where
$\cT(\l)$ is the set of component labels
$a\in\{1,\ldots,[(n+1)/2]\}$ for which $\tilde\a_a^2$ is allowed, by
the interlacing condition 
$\a\da\l$ from \nnn{BRANCH}, to take on more than one value; and
$$
s_u=\frac12(|\tilde\l|^2-|\tilde\m_u|^2).
$$
(The precise value of $\tilde c_u$ will become important below;
it is given in \cite{tbsw}, Theorem 5.2, and
in Theorem 6.1 below.)
By \cite{tbsw}, the cardinality of $\cT(\l)$ is $t(\l)$ (the same
as the dimension of the space generated by the 
various $\z_u(\x)^*\z_u(\x)$).

By \cite{ssgs}, Corollary 8.2, if $\l\lra\l$,
the constant $\tilde c_{\l\l}$ is positive,
and the eigenvalue of $D_{{\rm self}}$
on $\cV(\a;\l)$ is
$$
({\rm sgn}\,\a_{\ell+1})\sqrt{\eig(D_{{\rm self}}^2,\a)}
=({\rm sgn}\,\a_{\ell+1})
\sqrt{\tilde c_{\l\l}}\prod_{a\in\cT(\l)}|\tilde\a_a|.
$$
(Recall from Remark \ref{SelfG} 
that $D_{{\rm self}}$ is well-defined only up to 
an overall sign.)
Thus, since $\tilde\a_1=j+O(j^0)$,
\beq\label{dps2}
\dpsC_2(G_u^*G_u)(\acirc)=\tc_u\prod_{1<a\in\cT(\l)}
(\tilde\a_a^2-s_u^2),
\eeq
and if $\l\lra\l$,
\beq\label{selfspec}
\dpsC_1(D_{\rm self})=({\rm sgn}\,\a_{\ell+1})
\sqrt{\tilde c_{\l\l}}\prod_{1<a\in\cT(\l)}|\tilde\a_a|.
\eeq

If $\b\in\c(n-1)$, let 
\beq\label{barb}
\bar\b=\left\{\begin{array}{l}
\b\ {\rm when}\ n-1\ {\rm is\ odd}, \\
(\b_1,\ldots,-\b_\ell)\ {\rm when}\ n-1\ {\rm is\ even}.
\end{array}\right.
\eeq
Note that $\l\da\b\iff\l\da\bar\b$, and that there exist
$H(n-1)$-modules $\b$ with $\l\da\b$ and $\bar\b\ne\b$ if and only
if $n$ is odd and $\l_\ell\ne 0$.  This is exactly the case in which
there is a self-gradient on $\bbV(\l)$.
Note also that the
$\dpsC_2(G_u^*G_u)$ are constant on sets $\{\b,\bar\b\}$.
We claim that the $G_u^*G_u$ 
separate the various sets $\{\b,\bar\b\}$ in $B(\l)$, in the sense that
$$
\g\notin\{\b,\bar\b\}\;\Rightarrow\;\exists u:\,
\dpsC_2(G_u^*G_u)(\g)\ne\dpsC_2(G_u^*G_u)(\b).
$$
If we assume the contrary, then by \nnn{dps2}, the monic polynomials
\begin{equation}\label{monics}
\prod_{1<a\in\cT(\l)}(x^2-\tilde\b_a^2),\qquad
\prod_{1<a\in\cT(\l)}(x^2-\tilde\g_a^2)
\end{equation}
agree at the points $x=\pm s_u$ for all $u=1,\ldots,N(\l)$.
These polynomials have degree $2(t(\l)-1)$, which is either
$N(\l)-2$ or $N(\l)-1$.  There are $N(\l)$ labels $u$.  The $s_u$
comprise a set of cardinality $N(\l)$ {\em except} in the following
cases:
\begin{itemize}
\setlength{\itemsep}{-1ex}
\item $n$ is even and $\l_{\h-1}\ne 0=\l_\h\,$, or
\item $n$ is odd and $\l_\h\ne 0$, or
\item $\l_\h=\pm\tf\,$.
\end{itemize}
(Note that the last two cases overlap.)  In these exceptional cases,
the $s_u$ make up a set of cardinality $N(\l)-1$.
This shows that the two monic polynomials in \nnn{monics} agree.
Thus $(\tilde\g_a^2)_{1<a\in\cT(\l)}$ is a permutation of
$(\tilde\b_a^2)_{1<a\in\cT(\l)}$.  By strict dominance of $\tilde\b$
and $\tilde\g$ (\nnn{dominant} with all $\ge$ signs replaced by $>$
signs, since $(\r_{n+1})_a>|(\r_{n+1})_{a+1}|$), this can only be
the identity permutation, and the claim is proved. 

We now claim that for sufficiently large $k$, there are 
$2k$-homogeneous polynomials $P_\b$ in the $G_u^*G_u$, one for
each $\b\in B(\l)$, with $\dpsC_{2k}(P_\b)(\b)=
\dpsC_{2k}(P_\b)(\bar\b)=1$ and $\dpsC_{2k}(P_\b)(\g)=0$ for
$\g\notin\{\b,\bar\b\}$.  
Indeed, if 
$\dpsC_2(G_u^*G_u)$ separates $\{\b,\bar\b\}$ and $\{\g,\bar\g\}$,
say
$$
\dpsC_2(G_u^*G_u)(\b)\ne\dpsC_2(G_u^*G_u)(\g)=:C,
$$
then $Q_{\b\g}:=C\N^*\N-G_u^*G_u$ has
$$
\dpsC_2(G_u^*G_u)(\b)\ne 0=\dpsC_2(G_u^*G_u)(\g)=:C,
$$
Now let $P_\b$ be the composition, in some order, of the $Q_{\b\g}$
for the various $\g$ (one factor for each $\{\g,\bar\g\}$), and normalize.
(Note that the $Q_{\b\g}$ do not necessarily commute, but their realizations
on the sphere commute.  In particular, their leading symbols commute.)

We still need to separate $\b$ from $\bar\b$ when they are distinct.
By \nnn{selfspec}, this is accomplished by composing the operators
$P_\b$ with $D_{{\rm self}}$.  
($D_{{\rm self}}$ exists by the remarks following the definition
\nnn{barb}.
We may compose on either side; the
difference is a lower-order operator, again by the multiplicity free
nature of the $H(n+1)$-to-$H(n)$ branching rule applied to the
realizations of the operators on the sphere.)
After normalization, this gives us $(2k+1)$-homogeneous polynomial
operators $P'_\b$ with
$$
\dpsC_{2k+1}(P'_\b)(\b)=-\dpsC_{2k+1}(P'_\b)(\bar\b)=1,\qquad
\dpsC_{2k+1}(P'_\b)(\g)=0\ {\rm for}\ \g\notin
\{\b,\bar\b\}.
$$

The first statement of the theorem is now established.
For the second statement, we just note that the $P_\b$ already
separated points of $B(\l)$ in the case where there is no
self-gradient.
\end{proof}

Thus by dimension count and order parity, we have:

\begin{corollary}\label{zetaspan}
The restrictions to the unit $\x$-sphere of polynomials in the 
$\z_u(\x)^*\z_u(\x)$ generate $\cA_0(\l)$, and, when $\dself$ exists,
the restrictions of $\uself(\x)$ times such polynomials generate
$\cA_1(\l)$.  
If $n$ is even or $\l_\ell=0$, we have 
$\cA(\l)=\cA_0(\l)$.
\end{corollary}

The last conclusion of the corollary can actually be seen by elementary
means for $n$ even, since for each weight $m$ of the module $\l$,
the numbers $m_1+\cdots+m_\ell$ and $\l_1+\cdots+\l_\ell$ in 
$\frac12\bbZ$ have to agree
mod 2.

\begin{remark}\label{caution}
{\rm When
realizing
a given reduced symbol $\k(\x)$ using the $\z_u(\x)^*\z_u(\x)$ and 
possibly $\uself(\x)$, it is {\em a priori} possible that 
the minimal homogeneous degree in the $\z$
and $\Upsilon$ quantities may exceed the degree of $\k$.
It is probably reasonable to conjecture that this does {\em not} happen.
However, in this paper, we only establish this for 
bundles $\bbV(\l)$ with $N(\l)\le 4$
(Section \ref{low}).}
\end{remark}

\section{The relation between the local and global discrete symbols}

The estimate \nnn{ellest} shows that $\dpsC_p(D)$ depends only
on the leading symbol of $D$.  Thus the discrete symbol map
$\dpsC_p$ factors through $\tilde\s_p(\cG^p)$:
$$
\dpsC_p(D)=\dpsD_p(\tilde\s_p(D)),
$$
for some map $\dpsD_p$ on the space of restrictions to the
unit $\x$-sphere of $p$-homogeneous polynomials
in the $\z_u(\x)^*\z_u(\x)$ and, if applicable, $\uself(\x)$.
(Compare \nnn{factorA}.)
Let $\dpsD$ be the induced map on the space of all 
(not necessarily homogeneous) polynomials.  

\begin{theorem}\label{samesymb} The maps $\dpsA$ and $\dpsD$ carrying
$\cA(\l)$ to ${\rm maps}(B(\l),\bbC)$ are identical.
\end{theorem}

\begin{proof}{Proof} By 
Theorem \ref{gradspan}, it will be enough to show that $\dpsA$
and $\dpsD$ agree on the $\swsyu$ and, if applicable, $\uself$.
Thus it will be enough to show that 
$\dpsB_2$ and $\dpsC_2$ agree on the $\swu$ and, if applicable,
$\dpsB_1$ and $\dpsC_1$ agree on $\dself$.

Let $L_{ij}$ be the standard generators of $\gso(n+1)$.  By the 
spectral formula \nnn{TheSpectrum}, 
the action of the differential operator $G_u^*G_u$
on the highest weight vector of each $\gso(n+1)$-type
is given by the action, in the extension $S$
of the infinitesimal representation $\ind_{H(n)}^{H(n+1)}\l$
to the enveloping algebra of $\gso(n+1)$, of
$$
Q_u:=\tilde c_u
\prod_{a\in\cT(\l)}\left(\left(-\eye L_{2a-1,2a}+\frac{n+1}2-a\right)^2
-s_u^2\right).
$$
For a differential operator $D$ of order $p$ and a $\ci$ function $f$,
$$
\s_p(D)(x,(df)_x)\f_x=\lim_{t\to\infty}t^{-p}D(e^{\eye ft}\f),
$$
where $\f$ is any smooth extension of $\f_x\in\bbV_x$ to a section of 
$\bbV$.
By the branching rule \nnn{BRANCH}, the $H(n+1)$-types appearing in the
section space where $f$ lives, i.e.\ that over the trivial representation,
are the $(j,0,\ldots,0)$ for $j\in\bbN$.  
Choosing $f$ to depend on only one homogeneous coordinate, say 
$x^2$, is equivalent
to choosing $f$, or any function of $f$, to be the sum of highest weight
vectors.
Choosing $\f$ to be a highest weight vector as above, 
we get
$$
\begin{array}{l}
G_u^*G_u(e^{\eye tf}\f)=S(Q_u)(e^{\eye tf}\f) \\
{\ }=\left(\left(-\eye L_{12}+\frac{n-1}2\right)^2-s_u^2
\right)
\left(e^{\eye tf}\prod_{1<a\in\cT(\l)}
\left(\left(-\eye L_{2a-1,2a}+\frac{n+1}2-a\right)^2
-s_u^2\right)\f\right) \\
{\ }=\left(\left(-\eye L_{12}+\frac{n-1}2\right)^2-s_u^2
\right)
\left(e^{\eye tf}
\sum_{\b\in B(\l)}\dpsC(G_u^*G_u)(\b)\f_\b\right),
\end{array}
$$
where $\f_\b$ is the component of 
$\f$ in the direct sum of all $(\l_1+j,\b)$ summands in \nnn{sections}.
Now 
realize $S^n$ as $H(n+1)/H(n)$, where the Lie algebra of 
$H(n)$ is the $\gso(n)$ stabilizing $e_1$.  
At the identity coset {\bf o}, 
The right $H(n)$ covariance type of $\f_b$ {\em at} {\bf o} (but
not necessarily anywhere else) is the 
left $H(n)$ covariance type of $\f_b$, namely $\b$.
At {\bf o}, the covector $df$ naturally 
picks a multiple of the unit covector
$\x$ corresponding
to $e_2$.  

Substituting into the above, we have
that
$$
\s_2(G_u^*G_u)({\bf o},e_2)\f_x=
\sum_{\b\in B(\l)}\dpsC(G_u^*G_u)(\b)\left(\f_x)_\b\right),
$$
where $(\f_x)_\b$ is the projection of 
$\f_x\in\bbV_{{\bf o}}$
to the summand of right $H(n-1)$ covariance type $\b$.
On the other hand, by definition,
$$
\s_2(G_u^*G_u)({\bf o},e_2)\f_x=
\sum_{\b\in B(\l)}\dpsB(G_u^*G_u)(\b)\left(\f_x)_\b\right).
$$
By the generating property of Corollary \ref{zetaspan}, 
the transforms $\dpsA$ and $\dpsD$
agree on $\cA_0(\l)$, and thus on $\cA(\l)$ when there is no
self-gradient.  The self-gradient, when it exists, is handled by
an entirely similar argument, using
the action of the enveloping algebra element
$$
\sqrt{c_{\l\l}}\prod_{a=1}^{(n+1)/2}\left(\eye L_{2a-1,2a}
+\frac{n+1}2-a\right).
$$ 
\end{proof}

\begin{corollary}\label{ellcor} Let $D$ be natural differential operator
of order $p$
on some $\bbV(\l)$.
\newline
{\bf(a)} $D$ has spectral asymptotics on the sphere of
the form
$$
\eig(D_{S^n},\cV((\l_1+j,\b);\l))=F(D,\b)j^p+O(j^{p-1}),
$$
where the coefficient $F(D,\b)$ is the pointwise-determined 
eigenvalue of
$\s_p(D)(\x)$ on the summand of $\bbV(\l)_x$ transforming according
to the representation $\b$ of the $H(n-1)$ subgroup 
fixing $\x$ (on any $H(n)$ manifold, for any point $x$, at any
nonzero covector $\x$).
\newline
{\bf(b)} This natural differential operator is elliptic if and only
if its asymptotics on the sphere satisfy $F(D,\b)\ne 0$ for all
$\b\in B(\l)$.  It has positive definite leading symbol if and only if
$F(D,\b)>0$ for all $\b\in B(\l)$.
\newline
{\bf(c)} For some $k\ge 0$,
$(\N^*\N)^kD$
is a $(p+2k)$-homogeneous polynomial in the $\swu$ and, if applicable,
$\dself$, modulo operators of order at most $p+2k-1$.
\end{corollary}

\begin{remark}\label{misc}
{\rm The following remarks are clear from the above:
\begin{itemize}
\item The range of $\cA_0(\l)$ (resp.\ $\cA_1(\l)$) under $\dpsA$ 
consists of those functions on $B(\l)$ which are even (resp.\ odd)
under $\b\mapsto\bar\b$.
\end{itemize}
}
\end{remark}

\begin{remark}\label{whichp}
{\rm The {\em degree} of an element $\u$ of $\cA(\l)$ is the maximal
$p$ for which the $\cA^p(\l)$ component of $\u$ is nonzero.
It is clear that
$$
\deg\left((\z_{u_1}(\x)^*\z_{u_1}(\x)
\cdots\z_{u_{k+1}}(\x)^*\z_{u_{k+1}}(\x))\tilde{\ }\right)
\le 2+\deg\left((\z_{u_1}(\x)^*\z_{u_1}(\x)
\cdots\z_{u_k}(\x)^*\z_{u_k}(\x))\tilde{\ }\right)
$$
By the generating property of Corollary \ref{zetaspan}, 
the existence of nonzero, degree $p$
homogeneous polynomials (and thus monomials) guarantees the existence
of monomials of degrees $p-2,p-4,\ldots(0\,{\rm or}\,1)$.
Thus
\begin{itemize}
\item $\hom_{H(n)}(\tfs^{2k}\otimes\l,\l)\ne 0,\qquad k=0,\ldots,d_0(\l)/2$,
\item $\hom_{H(n)}(\tfs^{2k+1}\otimes\l,\l)\ne 0,\qquad k=0,\ldots,
(d_1(\l)-1)/2$.
\end{itemize}
}
\end{remark}

\section{Low $N(\l)$}\label{low}

There is an essential simplification of the above theory for bundles
$\bbV(\l)$ whose $N(\l)$ is at most $4$.  It is worthwhile to work
this out in detail because these are the bundles one is most likely to
meet ``in real life''.  For example, differential form bundles have
$N(\l)=3$ (or $N(\l)=2$, for the half middle-form bundles in even
dimensions).  The trace-free symmetric bundles $\bbV(p,0,\ldots,0)$
have $N(\l)=3$ for $p>0$.  Spinor bundles have $N(\l)=2$, and twistor
bundles have $N(\l)=4$.  Bundles of algebraic Weyl tensors have
$N(\l)=2$ (for $n=4$), $N(\l)=3$ (for $n=5$ and $n\ge 7$), or
$N(\l)=4$ (for $n=6$).  What we find for $N(\l)=3,4$ is the type of
information that is typically computed case by case, using some
explicit realization of the bundle and operator involved.  What we get
from the discrete leading symbol is a realization that all of these
computations are special cases of universal results, and in fact may
be obtained by substituting the weight parameter $\l$ into certain
universal formulas.

Suppose $N(\l)$ is 3 or 4.  Then $t(\l)$, the number of linearly
independent $\swu$, is 2.  This means that the restrictions of leading
symbols of $\N^*\N$ and any single $\swu$ (i.e., 1 and a single
$\swsyun$) generate $\cA_0(\l)$.  
(We just need to verify that these symbols are linearly independent;
this is clear from \nnn{dps2}.)
By finite dimensionality, the list
$$
1,\ \swsyun,\ (\swsyun)^2,\ ...,\ (\swsyun)^k,\ ...
$$
eventually becomes linearly dependent.  Thus there is a {\em minimal
polynomial} $M_{\l,u}(x)$ with
\beq\label{mnml}
\begin{array}{c}
f(\swsyun)=0,\ f\in\bbC[x]\ \iff\ f|M_{\l,u}, \\
\cA_0(\l)\cong\bbC[x]/(M_{\l,u}), \\
b_0(\l)=\deg M_{\l,u}.
\end{array}
\eeq
The minimal polynomial must also be the product of the {\em distinct}
$x-\dpsA(\swsyun)(\b)$ over all $\b\ua\l$.  Thus (just counting
degrees) each of the $b_0(\l)$
classes $\{\b,\bar\b\}$ for $\b\ua\l$ must take a different value under
$\dpsA(\swsyun)(\b)$.  Since $t(\l)=2$, there is
just one value $a_0$ of $a$ for which $\tilde\b_a^2$ is allowed more
than one value by the branching condition $\l\da\b$.  By \nnn{dps2}
and Theorem \ref{samesymb},
$$
\dpsA(\swsyun)(\b)=\tc_u(\tb_{a_0}^2-s_u^2).
$$
As a result,
\beq\label{mnml2}
M_{\l,u}(x)=\prod_{m=0}^{b_0(\l)-1}
\left(x-\tc_u\left(
\left(\l_{a_0}-m+\dfrac{n-1}2-a_0\right)^2-s_u^2\right)\right).
\eeq
The minimal polynomials $M_{\l,u}$ for different $u$ are closely related:
in terms of the $u$-independent data
\beq\label{mnml3}
h_m:=\l_{a_0}-m+\dfrac{n-1}2-a_0,\qquad
M_\l(x):=\prod_{m=0}^{b_0(\l)-1}(x-h_m^2),
\eeq
we have
$$
M_{\l,u}(x)=\tc_u^{b_0(\l)}M_\l(\tc_u^{-1}x+s_u^2).
$$
In particular, the set of roots of $M_{\l,u}$ is obtained from that
of $M_\l$ by an affine map on $\bbR$.
By \nnn{mnml2}, the fundamental projections on $\cA_0(\l)$ are
$$
\P_m(\eta):=\prod_{m'\ne m}\dfrac{\swsyun-\tc_u(h_{m'}^2-s_u^2)}
{\tc_u(h_m^2-s_u^2)-\tc_u(h_{m'}^2-s_u^2)}
=\tc_u^{-b_0(\l)+1}\prod_{m'\ne m}\dfrac{\swsyun-\tc_u(h_{m'}^2-s_u^2)}
{h_m^2-h_{m'}^2}\,.
$$
They are represented by the differential operators
\beq\label{prodcomp}
\cD_{m,u}:=\tc_u^{-(b_0(\l)+1)}
\prod_{m'\ne m}\dfrac{\swu-\tc_u(h_{m'}^2-s_u^2)\N^*\N}
{h_m^2-h_{m'}^2}\,,
\eeq
which have order at most $2(b_0(\l)-1)$.
Note that the projections $\P_m$ are independent of $u$; that is,
independent of which $\swsyun$ we have chosen as the preferred generator
for $\cA_0(\l)$.  The $\cD_{m,u}$ are in general not independent of
$u$; what one {\em can} say is that any $\cD_{m,u}-\cD_{m,v}$
has order at most $2(b_0(\l)-2)$.  (This order is strictly less
than $2(b_0(\l)-1)$, but by invariant theory, each term must
introduce a curvature, dropping the order by at least 2.)
In fact, the product in \nnn{prodcomp}, which is really a composition,
is sensitive to the ordering of the factors, as $G_u^*G_u$ and 
$\N^*\N$ commute only modulo second order operators.  Thus
$\cD_{m,u}$ is really only well-defined modulo operators of order 
at most $2(b_0(\l)-2)$.
 
Among other things, these results illuminate a typical
experience in computing with leading symbols on a bundle: eventually,
as the symbol order goes up, there are no new combinatorial
interactions of symbol and bundle indices.

Since $(\swsyu)^k$ can involve actions of at most symmetric $2k$-tensors,
we have in addition
\beq\label{est0}
d_0(\l)\le 2((\deg m_{\l,u})-1)=2(b_0(\l)-1)=
\left\{\begin{array}{l}
2[\l_\ell],\ n\ {\rm odd},\ a_0=\ell, \\
2(\l_{a_0}-|\l_{a_0+1}|)\ {\rm otherwise}.
\end{array}\right.
\eeq
If there is no self-gradient, $\cA_0(\l)$, $b_0(\l)$,
and $d_0(\l)$ can be
replaced by $\cA(\l)$, $b(\l)$, and $d(\l)$ respectively in the above
remarks.
The estimate \nnn{est0} becomes
$$
d(\l)\le 2(\l_{a_0}-|\l_{a_0+1}|),
$$
a clear improvement on Lemma \ref{Can'tAct} unless $a_0=1$ and $\l_{a_0+1}=0$.

Still in the case $N(\l)=3$ or 4, 
if there is a self-gradient, its restricted symbol together with 1
generate $\cA(\l)$.  There is a minimal polynomial $m_{\l,{\rm self}}(x)$
with
\beq\label{degm}
\begin{array}{l}
f(\uself)=0,\ f\in\bbC[x]\ \iff\ f|m_{\l,{\rm self}}, \\
\cA(\l)\cong\bbC[x]/(m_{\l,{\rm self}}), \\
b(\l)=\deg m_{\l,{\rm self}}.
\end{array}
\eeq
In fact, reasoning as above, if we let $u$ be the index corresponding
to the self-gradient (so that $\dself^2=\swu$), then
$$
\begin{array}{rl}
m_{\l,{\rm self}}(x)&=\prod_{m=0}^{b(\l)-1}\left(x-\sqrt{\tc_u}
(\l_{a_0}-m+\frac{n-1}2-a_0)\right) \\
&=\left\{\begin{array}{l}
\prod_{m=0}^{b_0(\l)-1}(x^2-\tc_uh_m^2),\ \ \l_\ell\in\frac12+\bbN, \\
x\prod_{m=0}^{b_0(\l)-2}(x^2-\tc_uh_m^2),\ \ \l_\ell\in\bbZ^+ 
\end{array}\right. \\
&=
\left\{\begin{array}{l}
M_{\l,u}(x^2)=\tc_u^{b_0(\l)}M_\l(\tc_u^{-1}x^2)
,\ \l_\ell\in\frac12+\bbN, \\
M_{\l,u}(x^2)/x=\tc_u^{b_0(\l)}M_\l(\tc_u^{-1}x^2)/x
,\ \l_\ell\in\bbZ.
\end{array}\right.
\end{array}
$$
Note that $s_u=0$ in this case.  Note that if $a_0=\ell$, the first
expression for the minimal polynomial simplifies:
$$
m_{\l,{\rm self}}(x)=\prod_{m=0}^{b(\l)-1}\left(x-\sqrt{\tc_u}
(\l_\ell-m)\right)\ \ {\rm if}\ a_0=\ell.
$$

The fundamental projections on $\cA(\l)$ are
$$
\P_m=\prod_{m'\ne m}\frac{\uself(\eta)-\sqrt{\tc_u}h_{m'}}
{\sqrt{\tc_u}h_m-\sqrt{\tc_u}h_{m'}}
=\tc_u^{(-b(\l)+1)/2}\prod_{m'\ne m}\frac{\uself(\eta)-\sqrt{\tc_u}h_{m'}}
{m'-m}
$$
where $h_m$ is defined as before, in this new range of $m$.
If $a_0=\ell$, then $h_{m'}$ is just $\l_\ell-m'$.
These projections generally have no operator representatives, since
they mix even and odd orders.

If $N(\l)=2$, then $\cA_0(\l)$ is generated (and thus spanned) by 1.  
If there is a self-gradient, $\cA_1(\l)$ is spanned by the odd
function of absolute value 1, and thus $\cA(\l)$ is generated and spanned
by this function and 1. 
(In fact, if $N(\l)=2$ and there is a self-gradient,
then $n$ is odd and $\l=(\frac12,\ldots,\frac12)$.  If
$N(\l)=2$ and there is no self-gradient, then $n$ is even and
$\l=(p,\ldots,p,\pm p)$ with $p\ne 0$.)

If $N(\l)=1$, then there can be no self-gradient, and $\cA(\l)$ is generated
and spanned by the function 1.  (In fact, the only $N(\l)=1$ case is
$\l=0$.)

Summing up the assertions made and proved above, we have

\begin{theorem}\label{gensym} {\bf(a)} If $n$ is odd, $\l_\ell\ne 0$,
and $N(\l)$ is $3$ or $4$, 
then
the principal symbol algebra $\cA(\l)$ is 
generated by 1 and the leading symbol of $\dself\,$.  
{\bf(b)} In all other
cases where $N(\l)$ is 3 or 4, $\cA(\l)$ is generated by
1 together with the leading symbol of any $G_u^*G_u\,$.
{\bf(c)} If $n$ is odd and $\l=(\tf,\ldots,\tf)$, then $N(\l)=2$, and
$\cA(\l)$
is generated by the leading symbol of the Dirac operator.
{\bf(d)} If $n$ is even and $\l=(p,\ldots,p,\pm p)$ with $p\ne 0$,
then $N(\l)=2$, and $\cA(\l)$ is generated by $1$.  {\bf(e)} If $\l=(0)$, then
$N(\l)=1$, and $\cA(\l)$ is generated by $1$.

Cases {\bf(a)}--{\bf(e)} exhaust all bundles with $N(\l)\le 4$.
\end{theorem}

On manifolds of constant curvature, we use this to get a statement
about generators of the algebra of invariant {\em operators} (as
opposed to symbols).  On a manifold of constant curvature, the
Weyl and trace-free Ricci tensors vanish, along with all covariant
derivatives of curvature; the only local tensorial invariants are
polynomials in the (constant) scalar curvature.

\begin{theorem}\label{genop} On a manifold of constant curvature,
the algebra $\cD(\l)$ of natural differential operators on 
$\bbV(\l)$, for $N(\l)\le 4$, 
is generated by the identity operator together with:
\begin{itemize}
\item $\N^*\N$ and $\dself$ in case {\bf(a)} above;
\item $\N^*\N$ and any single $G_u^*G_u$ in case {\bf(b)};
\item the Dirac operator in case {\bf(c)};
\item $\N^*\N$ in cases {\bf(d)} and {\bf(e)}.
\end{itemize}
\end{theorem}

\begin{proof}{Proof} 
By the constant curvature assumption, natural operators are polynomial
in the metric, the covariant derivative, the volume form, and (if applicable)
the fundamental tensor-spinor.
Let $D\in\cD(\l)$.  Theorem \ref{gensym} shows that 
there is an operator $P$ in the algebra generated by the 
putative generating set such that ${\rm ord}(D-P)<{\rm ord}(D)$.
Since $D-P$ is natural, the result follows by induction on the order.
\end{proof}

What goes wrong with the attempt to have something like Theorem
\ref{genop} in general is that the leading symbol of an operator like
$r^{ij}\N_i\N_j$ (where $r$ is the Ricci tensor) does not necessarily
induce an element of $\cA(\l)$.
 
As a corollary, we have:

\begin{corollary}\label{gencor} On the hyperboloid $G/K$, where
$G={\rm Spin}_0(n,1)$ and $K={\rm Spin}(n)$,
the $G$-invariant
differential operators on 
$\bbV(\l)$, for $N(\l)\le 4$, are generated by 
the operators listed in Theorem \ref{genop}.  If $\l$ is integral,
we may replace ``Spin'' by ``SO'' in the definitions of $G$ and $K$.
\end{corollary}

\section{Examples}\label{exs}

In the following, let $\eta$ be 
a vector on the unit sphere in the cotangent bundle.  
In doing explicit examples, it is helpful to know explicitly all the linear
relations among the various $\z_u(\x)^*\z_u(\x)$.  These are given
in \cite{tbsw}, Theorem 5.10:

\begin{theorem}\label{allWeit}
Given $\l\in\c(n)$, 
$$
\sum_{u=1}^{N(\l)}b_u\z_u(\x)^*\z_u(\x)=0
$$
if and only if
$$
\sum_{u=1}^{N(\l)}b_u\tilde c_{\l\s_u}s_u^{2j}=0,\qquad
j=0,1,\ldots,t(\l)-1.
$$
Here $\tilde c_{\l\s_u}=$
\begin{eqnarray}
&\dfrac{(-1)^{t(\l)+1}}{\displaystyle\prod_{{\tiny
\begin{array}{c}1\le v\le N(\l) \\ v\ne u\end{array}}}
(s_v-s_u)}\qquad &\text{if}\ N(\l)\ \text{is\ odd;}\label{c1} \\
&\dfrac{(-1)^{t(\l)+1}}
{2\displaystyle\prod_{u=1}^{N(\l)-2}(s_u+\tf)}\qquad
&\text{if\ }n\ \text{ is\ even,\ }\l_\h=0\ne\l_{\h-1}\,,\ \s_u=\l\pm e_\h\,;
\label{c2} \\
&\dfrac{(-1)^{t(\l)}\left(s_u+\tf\right)}
{\displaystyle
\prod_{{\tiny
\begin{array}{c} 1\le v\le N(\l) \\ v\ne u\end{array}}}
(s_v-s_u)}\qquad &\text{otherwise}.\label{c3}
\end{eqnarray}
\end{theorem}

The values of $\c_{\l\s_u}$ were given in \cite{tbsw}, Theorem 5.2.
(Recall from \nnn{TheSpectrum} that the $\tilde c_u$
appear in the spectral asymptotics of the $G_u^*G_u$ on the sphere.)

The parameter $t(\l)$, in addition to the roles it plays above, is
also
\beq\label{alteregot}
t(\l)=\dim\hom_{H(n)}((\tfs^0\oplus\tfs^2)\otimes\l,\l)
\eeq
by, for example, \cite{pspum}, p.57.

We shall adopt the convention of ordering the selection rule targets
in decreasing (lexicographical) order.

\begin{example}\label{ExSpinorOddD} 
{\rm Let $\bbV(\l)$ be (Spin$(n)$-isomorphic to) 
the spinor bundle $\S$ for $n\ge 3$ odd.
Then $\l=(\frac12,\ldots,\frac12)$ and 
$$
N(\l)=2,\ t(\l)=1,\ b_0(\l)=b_1(\l)=1.
$$
There are actions of $\tfs^0$ and (by the selection rule) $\tfs^1$ on $\l$.
These exhaust the total of $b(\l)=2$ linearly independent actions, so
$$
d_0(\l)=0,\ d_1(\l)=1.
$$
The $\tfs^1$ action is, in fact, Clifford multiplication.
The gradient targets are $\s=(\frac32,\frac12,\ldots,\frac12)$ and
$\l$ itself.  $\bbV(\s)$ is the {\em twistor bundle}, and may be
realized as spinor-one-forms $\f$ which are annihilated by interior
Clifford multiplication: $\g^\a\f_\a=0$.
The fundamental projections on $\cA(\l)$ are 
$$
\frac{\id\pm\eye\gamma(\eta)}2\,.
$$
Since these mix even and odd orders (as will always be the case 
when there is a self-gradient), there are no 
differential operators
representing these projections.  There will always be differential
operators representing the fundamental projections of $\cA_0(\l)$,
which in this case is just one-dimensional.
Since
\beq\label{spinorsc}
s_1=-n/2,\ s_2=0,\ \tc_1=(n-1)/n,\ \tc_2=1/n,
\eeq
$\dpsB_2(\dself^2)$ is the constant function $1/n$.  
Let $\dc$ be the Dirac operator.  Since 
$\dc^2$ has the same leading symbol as $\N^*\N$, we have
$\dc=\sqrt{n}\dself$.}
\end{example}

\begin{example}\label{ExSpinEven} 
{\rm Let $\bbV(\l)$ be the positive spinor bundle $\S_+$ for 
$n\ge 4$ even.  (The considerations for $\S_-$ are entirely
analogous.)  
Then $\l=(\frac12,\ldots,\frac12)$ and 
$$
N(\l)=2,\ t(\l)=1,\ b(\l)=b_0(\l)=1.
$$
Thus there are no actions of trace-free symmetric tensors beyond
that of $\tfs^0$, and
$$
d(\l)=0.
$$
The identity generates $\cA(\S)$; note that
there is no analogue of the Dirac leading symbol, since the
Dirac operator carries $\S_+$ to $\S_-\,$.
(The fact that $\tfs^1\otimes\S_+$ contains a
copy of $\S_-$ reflects the fact that Clifford multiplication
carries $\S_+$ to $\S_-$.)
As for normalizations, \nnn{spinorsc} is still good, so again
$\dc^2=G_2^*G_2$.}
\end{example}

\begin{example}\label{dfform} {\rm Let $\bbV(\l)$
be the differential form bundle $\L^k$ for $0<k<(n-2)/2$.
Then
$\l=(1_k)$, and the gradient targets
are $(2,1_{k-1})$, $(1_{k+1})$,
and $(1_{k-1})$.  There is no self-gradient, so $\cA(\l)=\cA_0(\l)$.
We have
$$
N(\l)=3,\ t(\l)=2,\ b(\l)=2.
$$
Combining this with \nnn{alteregot}, we deduce the existence of
a $\tfs^2$ action, which together with the obvious $\tfs^0$
action exhausts the possible $\tfs^p$ actions:
$$
d(\l)=2.
$$
(Lemma \ref{Can'tAct} already
implies that $d(\l)\le 2$.)
The fundamental projections are $\i(\eta)\e(\eta)$ and $\e(\eta)\i(\eta)$,
where $\e$ and $\i$ are exterior and interior multiplication; their
differential representatives are the familiar Hodge operators $\d d$
and $d\d$.}
\end{example}

\begin{example}\label{TfsExpl} 
{\rm Let $\bbV(\l)=\TFS^p$, and suppose that $p\ge 2$, $n\ge 5$.  
There is no self-gradient, so $\cA(\l)=\cA_0(\l)$.  We have $\l=(p)$ and
$$
N(\l)=3,\ t(\l)=2,\ b(\l)=p+1.
$$
There is an action of $\tfs^{2p}$ on $\tfs^p$, namely
$$
\f_{a_1\cdots\a_p}\mapsto\Y_{a_1\cdots\a_p}{}^{b_1\cdots b_p}
\f_{b_1\cdots b_p}\,.
$$
By Remark \ref{whichp},
there must therefore be actions of $\tfs^{2k}$ for $k=0,\ldots,p$.
This exhausts the available $b(\l)=p+1$ actions.  Thus we know that
$$
d(\l)=2p,
$$
and in fact we know $\dim_{{\rm SO}(n)}(\tfs^k\otimes\tfs^p,\tfs^p)$
for every $k$ and $p$.
The selection rule targets are $(p+1)$, $(p,1)$, and $(p-1)$, and we
have
$$
\begin{array}{lll}
s_1=-\frac12(n+2p-1), &
s_2=-\frac12(n-3), &
s_3=\frac12(n+2p-3), \\
\tilde c_1=-\dfrac1{(p+1)(n+2p-2)}\,, &
\tilde c_2=\dfrac1{(p+1)(n+p-3)}\,, &
\tilde c_3=-\dfrac1{(n+2p-2)(n+p-3)}\,.
\end{array}
$$
By Theorem \ref{allWeit}, the single linear relation among the 
$\z_u(\x)^*\z_u(\x)$ has coefficients $(b_u)$ (in the notation of
the theorem), where $(b_u)$ is the unique (up to constant multiple)
solution of the system
$$
\begin{array}{rl}
0=&\tilde c_1b_1+\tilde c_2b_2+\tilde c_3b_3 \\
=&\tilde c_1s_1^2b_1+\tilde c_2s_2^2b_2+\tilde c_3s_3^2b_3\,.
\end{array}
$$
Thus the linear relation is
$$
-p\swsyT+\swsyH+(n+p-2)\swsyE=0.
$$
This allows us to write everything in terms 
$|\xi|^2$ and a single $\swsyu$, say $\swsyT$:
$$
\begin{array}{rl}
\swsyH=&\dfrac1{n+p-3}\left\{(n+p-2)|\xi|^2-(n+2p-2)\swsyT\right\}, \\
\swsyE=&\dfrac1{n+p-3}\left\{-|\xi|^2+(p+1)\swsyT\right\}
\end{array}
$$

Now consider the discrete leading symbols of the above operators.
To avoid trivialities, we exclude the case $p=0$.
$B(\l)$ is the $(p+1)$-point space of all
$(q)\in\c(n-1)$ with $0\le q\le p$.
By \nnn{mnml3} with $m=p-q$, 
$$
h_m=q+\dfrac{n-3}2,
$$
and the minimal polynomial of $\swsyT$ is
$$
\begin{array}{rl}
M_{\l,1}(x)&=
\prod_{q=0}^p\left(x+\dfrac1{(p+1)(n+2p-2)}(h_m-\frac12(n+2p-1))
(h_m+\frac12(n+2p-1))\right) \\
&=\prod_{q=0}^p\left(x-\dfrac{(p-q+1)(q+p+n-2)}{(p+1)(n+2p-2)}\right).
\end{array}
$$

To see what is going on tensorially, we first need a formula for $G_1$.
If $\y_{a_0\cdots a_p}$
is a section of $T^*\otimes\TFS^p$, then its projection 
to $\TFS^{p+1}$
is
$$
(P\y)_{a_0\cdots a_p}:=
\frac1{p+1}\sum_{s=0}^p\y_{a_sa_0\cdots\hat a_s\cdots a_p}
-\a\sum_{s<t}g_{a_sa_t}\y^b{}_{ba_0\cdots\hat a_s\cdots\hat a_t\cdots a_p}\,,
$$
where the number $\a$ is determined by the condition that the $a_0a_1$
metric trace (and thus every other trace, by symmetry)
vanishes.  
A short calculation gives
$$
\a=\frac2{(k+1)(n+2k-2)}\,.
$$
For example, if $p=2$, then 
$$
(P\y)_{a_0a_1a_2}=\dfrac13(\y_{a_0a_1a_2}
+\y_{a_1a_0a_2}+\y_{a_2a_0a_1})
-\dfrac2{3(n+2)}(g_{a_0a_1}\y^b{}_{ba_2}+
g_{a_0a_2}\y^b{}_{ba_1}+
g_{a_1a_2}\y^b{}_{ba_0}).
$$
$G_1\f$ is just 
$P(\N\f)$, 
and $\z_1(\x)\f$ is just $P(\x\otimes\f)$; each
may be expanded according to the formula above.
Furthermore, since $G_1^*G_1=\N^*G_1\,$, we have
$$
(G_1^*G_1\f)_{a_1\cdots a_p}=
-\dfrac1{p+1}\N^{a_0}\left(
\sum_{s=0}^p\N_{a_s}\f_{a_0\cdots\hat a_s\cdots a_p}
-\dfrac2{n+2p-2}
\sum_{s<t}g_{a_sa_t}\N^b\f_{ba_0\cdots\hat a_s\cdots\hat a_t\cdots a_p}
\right).
$$
For example, if $p=2$, 
\beq\label{preS}
\bear{l}
(G_1^*G_1\f)_{ab}=-\dfrac13\N^c\left(
\N_c\f_{ab}+\N_a\f_{bc}+\N_b\f_{ac}\right. \\
\left.\qquad-\dfrac2{n+2}(g_{ca}\N^d\f_{db}
+g_{cb}\N^d\f_{da}+g_{ab}\N^d\f_{dc})\right).
\eear
\eeq

Let us concentrate on the example $p=2$ for a moment.
The minimal polynomial with respect to the gradient target 
$\s_1=(3)$ is
\beq\label{tfs2ml1}
M_{\l,1}(x)=\left(x-\dfrac{n}{n+2}\right)
\left(x-\dfrac{2(n+1)}{3(n+2)}\right)
\left(x-\dfrac13\right).
\eeq
The fundamental projections are
$$
\begin{array}{rl}
\P_0&=\dfrac{\left(\swsyT-\frac{2(n+1)}{3(n+2)}\right)
\left(\swsyT-\frac13\right)}
{\left(\frac{n}{n+2}-\frac{2(n+1)}{3(n+2)}\right)
\left(\frac{n}{n+2}-\frac13\right)} \\
&=\dfrac{9(n+2)^2}{2(3n-2)(n-1)}
\left(\swsyT-\frac{2(n+1)}{3(n+2)}\right)
\left(\swsyT-\frac13\right), \\
\P_1&=\dfrac{\left(\swsyT-\frac{n}{n+2}\right)
\left(\swsyT-\frac13\right)}
{\left(\frac{2(n+1)}{3(n+2)}-\frac{n}{n+2}\right)
\left(\frac{2(n+1)}{3(n+2)}-\frac13\right)} \\
&=\dfrac{9(n+2)^2}{n(3n-2)}
\left(\swsyT-\frac{n}{n+2}\right)
\left(\swsyT-\frac13\right), \\
\P_2&=\dfrac{\left(\swsyT-\frac{n}{n+2}\right)
\left(\swsyT-\frac{2(n+1)}{3(n+2)}\right)}
{\left(\frac13-\frac{n}{n+2}\right)
\left(\frac13-\frac{2(n+1)}{3(n+2)}\right)} \\
&=-\dfrac{9(n+2)^2}{2n(n-1)}
\left(\swsyT-\frac{n}{n+2}\right)
\left(\swsyT-\frac{2(n+1)}{3(n+2)}\right),
\end{array}
$$
and $\P_q$ is the projection onto the $(q)\in\c(n-1)$ summand in the 
decomposition of $(p)\in\c(n)$ under the SO$(n-1)$ subgroup fixing
$\eta$.  

Taking an explicit tensorial viewpoint and proceeding from scratch
in this example, 
there are three independent combinatorial interactions
of $\x$ and $\f\in\tfs^2$, namely
$$
\begin{array}{rl}
X_0(\x,\f):&=\f, \\
X_2(\x,\f)_{\a\b}:&=\eta^\l\eta_{(\a}\f_{\b)\l}-\frac1{n}g_{\a\b}
\eta^\l\eta^\m\f_{\l\m}, \\
X_4(\x,\f)_{\a\b}:&=\eta_\a\eta_\b\eta^\l\eta^\m\f_{\l\m}
-\frac1{n}g_{\a\b}\eta^\l\eta^\m\f_{\l\m}.
\end{array}
$$
(Recall that $\eta_\a\eta^\a=1$.)  
In fact, these formulas make explicit the actions of $\tfs^0$,
$\tfs^2$, and $\tfs^4$ on $\tfs^2$.
An alternative basis 
consists of the identity $X_0$ together with 
$$
\begin{array}{rl}
\z_1(\eta)^*\z_1(\eta)&=\dfrac13X_0+\dfrac{2n}{3(n+2)}X_2, \\
(\z_1(\eta)^*\z_1(\eta))^2&=\dfrac19X_0+\dfrac{2n(3n+4)}{9(n+2)^2}X_2
+\dfrac{2n(n-2)}{9(n+2)^2}X_4.
\end{array}
$$ 
The fact that we have exhausted the combinatorial possibilities means
that the cube of $\z_1(\eta)^*\z_1(\eta)$ will be a linear combination
of previous powers, and indeed, 
\beq\label{mnmltfs2}
\begin{array}{rl}
(\z_1(\eta)^*\z_1(\eta))^3&=\dfrac1{27}X_0
+\dfrac{2n(7n^2+18n+12)}{27(n+2)^3}X_2
+\dfrac{4n(n-2)(3n+2)}{27(n+2)^3}X_4 \\
&=\dfrac{2n(n+1)}{9(n+2)^2}-\dfrac{11n^2+18n+4}{9(n+2)^2}
\z_1(\eta)^*\z_1(\eta)+\dfrac{2(3n+2)}{3(n+2)}
(\z_1(\eta)^*\z_1(\eta))^2.
\end{array}
\eeq
But the difference of the extreme left and right sides of 
\nnn{mnmltfs2} is exactly the minimal polynomial of 
\nnn{tfs2ml1}, applied to $\z_1(\eta)^*\z_1(\eta)$.
That is, substitution into our general machinery checks with
the result of naive calculation.
}
\end{example}

\begin{example}\label{WlTensors}
{\rm An interesting and potentially useful example is the 
bundle $\cW$ of {\em algebraic Weyl tensors}; i.e.\ totally trace-free
tensors with the symmetries
$$
Y_{\a\b\l\m}=Y_{\l\m\a\b}=-Y_{\a\b\m\l}=-Y_{\a\l\m\b}-Y_{\a\m\b\l}.
$$
If $n\ge 7$, these are a realization of $\bbV(2,2,0,\ldots,0)$
(see \cite{strich}), a bundle with no self-gradient, and selection
rule targets 
$$
\s_1=(3,2,0,\ldots,0),\ \s_2=(2,2,1,0,\ldots,0),\ 
\s_3=(2,1,0,\ldots,0).
$$
Thus
$$
N(\l)=3,\ t(\l)=2,\ b(\l)=3.
$$
There is one action of $\tfs^0$, and there is $t(\l)-1=1$ action
of $\tfs^2$.  
(Recall that in general, there are $t(\l)$ actions of $\tfs^0\oplus\tfs^2$ on 
any given $\l$, by \nnn{alteregot}.)
If $n$ is even, the only other possible action is
by $\tfs^4$, by parity considerations and Lemma \ref{Can'tAct}.
The tensorial formula for this must continue to odd dimensions,
so we have
$$
\dim\hom_{\so(n)}(\tfs^p\otimes\cW,\cW)=\left\{
\begin{array}{ll}
1,\ &p=0,2,4, \\
0\ &{\rm otherwise}.
\end{array}\right.
$$
Formulas for these actions will in fact emerge from the 
minimal polynomial calculations, much as in the previous
examples.

We choose to compute the minimal polynomial and projections from
the viewpoint of the third selection rule target $(2,1,0,\ldots,0)$.
Straightforward computation yields 
$$
\begin{array}{l}
s_3=\dfrac{n-1}2,\ 
\tc_{3}=-\dfrac1{(n+1)(n-3)}\,, \\
h_m=\dfrac{n-1}2-m,\qquad m=0,1,2.
\end{array}
$$
As a result the minimal polynomial of $\z_3(\eta)^*\z_3(\eta)$
is
\beq\label{xmnml}
M_{\l,3}(x)=\prod_{m=0}^2\left(x-\dfrac{m(n-1-m)}{(n+1)(n-3)}\right)
=x\left(x-\dfrac{n-2}{(n+1)(n-3)}\right)
\left(x-\dfrac{2}{n+1}\right),
\eeq
and the fundamental projections on $\cA(\l)$ are
$$
\begin{array}{rl}
\P_0&=\dfrac{\left(\z_3(\eta)^*\z_3(\eta)-\frac{n-2}{(n+1)(n-3)}\right)
\left(\z_3(\eta)^*\z_3(\eta)-\frac{2}{n+1}\right)}
{\left(-\frac{n-2}{(n+1)(n-3)}\right)
\left(-\frac{2}{n+1}\right)} \\
&=\dfrac{(n-3)(n+1)^2}{2(n-2)}(\z_3(\eta)^*\z_3(\eta))^2
-\dfrac{(n+1)(3n-8)}{2(n-2)}\z_3(\eta)^*\z_3(\eta)+1, \\
\P_1&=\dfrac{\z_3(\eta)^*\z_3(\eta)
\left(\z_3(\eta)^*\z_3(\eta)-\frac{2}{n+1}\right)}
{\frac{n-2}{(n+1)(n-3)}\left(\frac{n-2}{(n+1)(n-3)}-\frac{2}{n+1}
\right)} \\
&=\dfrac{(n+1)(n-3)^2}{(n-4)(n-2)}\left\{-(n+1)(\z_3(\eta)^*\z_3(\eta))^2
+2\z_3(\eta)^*\z_3(\eta)\right\}, \\
\P_2&=\dfrac{\z_3(\eta)^*\z_3(\eta)\left(\z_3(\eta)^*\z_3(\eta)-
\frac{n-2}{(n+1)(n-3)}\right)}
{\frac{2}{n+1}\left(\frac{2}{n+1}-\frac{n-2}{(n+1)(n-3)}\right)} \\
&=\dfrac{n+1}{2(n-4)}\left\{(n-3)(n+1)(\z_3(\eta)^*\z_3(\eta))^2
-(n-2)\z_3(\eta)^*\z_3(\eta)\right\}.
\end{array}
$$
(Here $\P_m$ is the projection on the branch $(2,2-m)$.)

To see what is going on tensorially, note that 
the symbol $\z_3(\eta)^*\z_3(\eta)$ is closely related to 
the symbol $\z'_1(\eta)$ of the top gradient $\bbV(2,1)\to\bbV(2,2)$:
$$
\z_3(\eta)^*\z_3(\eta)=C\cdot\z'_1(\eta)\z'_1(\eta)^*
$$
for some universal constant $C$.  This constant is easily evaluated
by the discrete principal symbol.
In fact, more generally, by \nnn{TheSpectrum},
$$
c_{\s\l}\dpsA(\z_{\l\s}(\eta)^*\z_{\l\s}(\eta))=c_{\l\s}
\dpsA(\z_{\s\l}(\eta)\z_{\s\l}(\eta)^*).
$$
In particular, the quotient $c_{\l\s}/c_{\s\l}$ can be evaluated by 
computing at any $\b\in\c(n-1)$ having $\b\ua\l$ and $\b\ua\s$.
Note that $c$ cannot be replaced by $\tc$ in this statement, 
since $\cT(\l)$ and $\cT(\s)$ need not be the same.  In fact, by
\nnn{TheSpectrum},
$$
\dfrac{\tc_{\l\s_u}}{\tc_{\s_u\l}}=
\dfrac{c_{\l\s_u}\prod_{a\in\cT_{\s_u}\setminus\cT_\l}(\ta_a^2-s_u^2)}
{c_{\s_u\l}\prod_{a\in\cT_{\l}\setminus\cT_{\s_u}}(\ta_a^2-s_u^2)}\,.
$$
In the present situation, with $\l=(2,2)$
and $\s_u=(2,1)$, we have
$$
\dfrac{\tc_{\l\s_u}}{\tc_{\s_u\l}}=
\dfrac{c_{\l\s_u}}{c_{\s_u\l}}(\ta_2^2-s_u^2),
$$
where $\ta_2^2=\left(\frac{n+1}2\right)^2$ is the only admissible
value of $\ta_2^2$, so
$$
\begin{array}{rl}
\dfrac{c_{\l\s_u}}{c_{\s_u\l}}&=\dfrac{4(n-2)}{(n+1)(n-3)}\,, \\
\dpsA(\z_{\l\s_u}(\eta)^*\z_{\l\s_u}(\eta))&=\dfrac{4(n-2)}{(n+1)(n-3)}
\dpsA(\z_{\s_u\l}(\eta)\z_{\s_u\l}(\eta)^*).
\end{array}
$$

One realization of $\bbV(2,1)$ is as the bundle
$\cH$ of 3-tensors that are totally trace free, 
antisymmetric in their last 2 indices, and Bianchi-like in all 3 indices.
(For a different realization, replace ``antisymmetric'' with ``symmetric''
in the last sentence.)  The natural operator $D$ defined by
$$
\begin{array}{rl}
(D\y)_{ijkl}:&=\dfrac14(\N_i\y_{jkl}-\N_j\y_{ikl}+\N_k\y_{lij}-\N_l\y_{kij})
\\
&\qquad-\dfrac1{4(n-2)}\N^m\left\{g_{ik}(\y_{jml}+\y_{lmj})
-g_{jk}(\y_{iml}+\y_{lmi})\right. \\
&\qquad\left.-g_{il}(\y_{jmk}+\y_{kmj})
+g_{jl}(\y_{imk}+\y_{kmi})\right\}.
\end{array}
$$
carries $\cH$ to $\cW$; thus it is a constant multiple of 
the corresponding realization of $G_{\s_3\l}$.
Writing $D\y=Q\N\y$, where $Q$ is a bundle map on $T^*M\otimes\cH$,
one finds $Q^2=Q$, so we have the correct normalization, and $D$ is 
exactly the realization of $G_{\s_3\l}$.  Since 
$$
(D^*\Y)_{ijk}=-\N^l\Y_{lijk},
$$
we have (in these realizations)
\beq\label{weylgg}
\begin{array}{rl}
(G_3^*G_3\Y)_{ijkl}
&=\dfrac{4(n-2)}{(n+1)(n-3)}(DD^*\Y)_{ijkl} \\
&=-\dfrac{n-2}{(n+1)(n-3)}(\N_i\N^p\Y_{pjkl}-\N_j\N^p\Y_{pikl}
+\N_k\N^p\Y_{plij}-\N_l\N^p\Y_{pkij})
\\
&\qquad+\dfrac1{(n+1)(n-3)}\N^m\N^p\left\{g_{ik}(\Y_{pjml}+\Y_{plmj})
-g_{jk}(\Y_{piml}+\Y_{plmi})\right. \\
&\qquad\left.-g_{il}(\Y_{pjmk}+\Y_{pkmj})
+g_{jl}(\Y_{pimk}+\Y_{pkmi})\right\}.
\end{array}
\eeq
$\z_3(\eta)^*\z_3(\eta)$ is obtained, of course, by replacing each
$\N_i\N_j$ with $-\eta_i\eta_j$ in the last expression:
$$
\begin{array}{rl}
(\z_3(\eta)^*\z_3(\eta)\Y)_{ijkl}
&=\dfrac{n-2}{(n+1)(n-3)}(\eta_i\eta^p\Y_{pjkl}-\eta_j\eta^p\Y_{pikl}
+\eta_k\eta^p\Y_{plij}-\eta_l\eta^p\Y_{pkij}) \\
&-\dfrac2{(n+1)(n-3)}\eta^m\eta^p\left\{g_{ik}\Y_{pjml}
-g_{jk}\Y_{piml}
-g_{il}\Y_{pjmk}
+g_{jl}\Y_{pimk}\right\}.
\end{array}
$$
To get 
an independent calculation of the minimal polynomial of 
$\z_3(\eta)^*\z_3(\eta)$ and of the fundamental projections, 
we just need to use \nnn{weylgg} to compute the powers of 
$\z_3(\eta)^*\z_3(\eta)$.
The results are
$$
\begin{array}{l}
((\z_3(\eta)^*\z_3(\eta))^2\Y)_{ijkl}=
\dfrac{(n-2)^2}{(n-3)^2(n+1)^2}\left\{\eta_i\eta^p\Y_{pjkl}
-\eta_j\eta^p\Y_{pikl}+\eta_k\eta^p\Y_{plij}-\eta_l\eta^p\Y_{pkij}\right\} \\
\qquad-\dfrac4{(n-3)(n+1)^2}\eta^m\eta^p\left\{g_{ik}\Y_{pjml}
-g_{jk}\Y_{piml}
-g_{il}\Y_{pjmk}
+g_{jl}\Y_{pimk}\right\} \\
\qquad+\dfrac{2(n-4)(n-2)}{(n-3)^2(n+1)^2}\eta^m\eta^p\left\{
\eta_j\eta_l\Y_{imkp}-\eta_j\eta_k\Y_{imlp}-\eta_i\eta_l\Y_{jmkp}
+\eta_i\eta_k\Y_{jmlp}\right\},
\end{array}
$$
and 
$$
\begin{array}{l}
((\z_3(\eta)^*\z_3(\eta))^3\Y)_{ijkl}=
\dfrac{(n-2)^3}{(n-3)^3(n+1)^3}\left\{\eta_i\eta^p\Y_{pjkl}
-\eta_j\eta^p\Y_{pikl}+\eta_k\eta^p\Y_{plij}-\eta_l\eta^p\Y_{pkij}\right\} \\
\qquad -\dfrac8{(n-3)(n+1)^3}\eta^m\eta^p\left\{g_{ik}\Y_{pjml}
-g_{jk}\Y_{piml}
-g_{il}\Y_{pjmk}
+g_{jl}\Y_{pimk}\right\} \\
\qquad+\dfrac{2(n-4)(n-2)(3n-8)}{(n-3)^3(n+1)^3}\eta^m\eta^p\left\{
\eta_j\eta_l\Y_{imkp}-\eta_j\eta_k\Y_{imlp}-\eta_i\eta_l\Y_{jmkp}
+\eta_i\eta_k\Y_{jmlp}\right\}.
\end{array}
$$
From this, one finds that for $A=\z_3(\eta)^*\z_3(\eta)$,
the only linear relation among 
$A^3$, $A^2$, $A$, and $1$ is, up to a constant multiple,
\beq\label{Amnml}
A^3-\dfrac{3n-8}{(n+1)(n-3)}A^2+\dfrac{2(n-2)}{(n+1)^2(n-3)}A=0.
\eeq
But the left side of \nnn{Amnml} factors to
$$
A\left(A-\dfrac{n-2}{(n+1)(n-3)}\right)
\left(A-\dfrac{2}{n+1}\right),
$$
exactly the minimal polynomial $M_{\l,3}(A)$ determined 
in \nnn{xmnml} from abstract considerations.
}
\end{example}

\begin{example}\label{oddtwistor}
{\rm A good example of a bundle with a self-gradient is 
the {\em twistor bundle} $\bbV(\l)$
for $\l:=(\frac32,\frac12,\ldots,\frac12)$
in odd dimensions.  
This bundle is realized in spinor-one-forms $\f$ with $\g^i\f_i=0$.
Suppose $n\ge 5$; then the selection rule targets are 
$$
\s_1=\l+e_1,\ \s_2=\l+e_2,\ 
\s_3=\l,\ \s_4=\l-e_1,
$$
Thus
\beq\label{twistdata}
N(\l)=4,\qquad t(\l)=2,\qquad b(\l)=4,\qquad b_0(\l)=2.
\eeq
The self-gradient is sometimes called the {\em Rarita-Schwinger
operator}; normalized to have square $G_3^*G_3$, it is
\beq\label{nmlrita}
(\cS\f)_i=\left(\dfrac{n}{(n+2)(n-2)}\right)^{1/2}
\left\{\g^j\N_j\f_i-\frac2{n}\g_i\N^j\f_j\right\}.
\eeq
in the twistor realization.  We denote the leading symbol of $\cS$
by $\Upsilon$.
$\Upsilon(\eta)^2$ will have a minimal polynomial $M_{\l,3}(x)$
of degree 
$2$ by \nnn{mnml}, and $\Upsilon(\eta)$ will have a minimal polynomial
$m_\l(x)$ of degree 4 by \nnn{degm}.  
Counting degrees, we conclude that
$$
m_\l(x)=M_{\l,3}(x^2).
$$
In particular, $m_\l$ is an even polynomial.
It is clear that there is one $\tfs^0$ action 
and one $\tfs^1$ action on
$\l$.  As a result, by \nnn{twistdata}, there is also 
one $\tfs^2$ action and one $\tfs^3$ action, and these exhaust
the possible $\tfs^p$ actions.  That is,
$$
\dim\hom_{\spin(n)}(\tfs^p\otimes\l,\l)=\left\{\begin{array}{l}
1,\ p=0,1,2,3, \\
0,\ p>3.\end{array}\right.
$$

Since
\beq\label{cRita}
\begin{array}{rl}
\tc_3&=\dfrac{4}{n(n+2)(n-2)}\,, \\
h_m&=\dfrac{n}{2}-m,\qquad m=0,1,
\end{array}
\eeq
our minimal polynomials and fundamental projections on $\cA_0(\l)$
are
\beq\label{twistproj}
\begin{array}{rl}
M_{\l,3}(x)&=\left(x-\dfrac{n}{(n+2)(n-2)}\right)
\left(x-\dfrac{n-2}{n(n+2)}\right), \\
\P_0&=\dfrac{\Ups(\eta)^2-\frac{n-2}{n(n+2)}}
{\frac{n}{(n+2)(n-2)}-\frac{n-2}{n(n+2)}}
=\dfrac{(n-2)\{n(n+2)\Ups(\eta)^2-(n-2)\}}{4(n-1)}\,, \\ 
\P_1&=\dfrac{\Ups(\eta)^2-\frac{n}{(n+2)(n-2)}}
{\frac{n-2}{n(n+2)}-\frac{n}{(n+2)(n-2)}}=-\dfrac{n\{(n+2)(n-2)
\Ups(\eta)^2-n\}}{4(n-1)}\,.
\end{array}
\eeq
$\P_m$ is the projection onto the $(\frac32-m,\frac12,\ldots,\frac12,
\pm\frac12)$ branches; $\P_0(\eta)$ and $\P_1(\eta)$
are represented by the second-order operators
$$
\dfrac{(n-2)\{n(n+2)\cS^2-(n-2)\N^*\N\}}{4(n-1)}\,,\qquad
-\dfrac{n\{(n+2)(n-2)
\cS^2-n\N^*\N\}}{4(n-1)}\,.
$$

To see what is going on tensorially, consider the (unnormalized)
leading symbol
of the Rarita-Schwinger operator,
$$
(\r(\eye\eta)\f)_i=\eye\left(\g^j\eta_j\f_i-\frac2{n}\g_i\eta^j\f_j
\right).
$$
We have
$$
\begin{array}{rl}
(\r(\eye\eta)^2\f)_i&=\f_i+\dfrac{4(n-1)}{n^2}\eta_i\eta^j\f_j
+\dfrac4{n^2}\a_{ik}\eta^j\eta^k\f_j, \\
(\r(\eye\eta)^3\f)_i&=\eye\left(\g^j\eta_j\f_i
+\dfrac{4(n-2)}{n^2}\g^k\eta_i\eta^j\eta_k\f_j
-\dfrac{2(n^2-2n+4)}{n^3}\g_i\eta^j\f_j\right), \\
(\r(\eye\eta)^4\f)_i&=\f_i+\dfrac{8(n-1)(n^2-2n+2)}{n^4}\eta_i\eta^j\f_j
+\dfrac{8(n^2-2n+2)}{n^4}\a_{ik}\eta^j\eta^k\f_j,
\end{array}
$$
where $\a$ is the antisymmetric Clifford symbol,
$$
\a^{ij}=\frac12(\g^i\g^j-\g^j\g^i).
$$
It is clear that these expressions exhaust the combinatorial
possibilities (subject to the Clifford relations, the twistor
condition, and $|\eta|^2=1$), that the 
$\r(\eye\eta)^p\f$ for $p=0,1,2,3$ are linearly independent, and
that
$$
\r(\eye\eta)^4=\dfrac{2(n^2-2n+2)}{n^2}\r(\eye\eta)^2-\dfrac{(n-2)^2}
{n^2}\,.
$$
That is, $\r(\eye\eta)^2$ satisfies the polynomial
$$
x^2-\dfrac{2(n^2-2n+2)}{n^2}x+\dfrac{(n-2)^2}{n^2}.
$$
The normalized symbol (from \nnn{nmlrita}) thus satisfies
$$
\dfrac{(n+2)^2(n-2)^2}{n^2}x^2-\dfrac{2(n+2)(n-2)(n^2-2n+2)}{n^3}x
+\dfrac{(n-2)^2}{n^2}.
$$
But this is exactly 
$$
\left(\dfrac{(n+2)(n-2)}{n}\right)^2M_{\l,3}(x),
$$
for the  $M_{\l,3}(x)$ predicted by \nnn{twistproj}.
}
\end{example}

\section{Applications} 

\subsection{Computation of Green's functions}

The discrete leading symbol allows us to compute Green's functions, or
fundamental solutions, of natural elliptic differential operators $D$ 
with nonscalar leading symbol, as follows.  
Suppose $D$ acts on sections of $\bbV(\l)$.
By ellipticity and 
Corollary \ref{ellcor}{\bf(b)}, $\dpsB$ is a nonzero function on 
$B(\l)$.  Let $\P_i$ be the fundamental projection corresponding to
the branch $\b_i\,$, and let $P_i$ be a natural differential 
operator with leading symbol $\P_i\,$.  Then
$$
E:=\sum_{i=1}^{b(\l)}\dfrac1{\dpsB(\b_i)}P_i
$$
is a natural differential operator for which the function
$\dpsB(DE)$ is identically 1; that is,
$$
DE=\Delta^k+{\rm(lower\ order)}
$$ 
for some $k$.  The computation of a fundamental solution for $D$,
that is, a distribution $G(x,y)$ for which $D_xG(x,y)=\delta_y(x)
{\rm Id}_{\bbV(\l)}$,
is now reduced to a similar computation for the operator $DE$,
whose leading symbol is less exotic.  For if $H$ is a fundamental 
solution for $DE$, then
$$
D_x(E_xH(x,y))=
(DE)_xH(x,y)=\delta_y(x){\rm Id}_{\bbV(\l)}\,.
$$
That is, $E_xH(x,y)$ is a fundamental solution for $D$.

The problem of computating asymptotic expansions and parametrices for
operators with principal part $\D^k$ is considerably more
straightforward than the same problem for operators with arbitrary
natural principal part.  With the fundamental projections in hand, we
have an {\em effective} procedure for getting {\em quasi-inverses} $E$
for natural elliptic $D$, and thus for effecting this reduction to the
case of scalar leading symbol.

The corresponding problem for the heat operator $\exp(-tD)$ is not
so greatly simplified by the computation of a quasi-inverse.
However, as shown in \cite{ab1}, it may also be attacked using
the $\P_i\,$.

\subsection{Conformally covariant operators}

The discrete spectral calculus 
allows a computation of the principal part of any conformally
covariant operator on sections of $\bbV(\l)$.  As a result, it also
gives a complete formula for any such operator in the conformally
flat case.

To describe this, we need a short summary of the state of knowledge
about conformally covariant operators.  All conformally covariant
differential operators in the conformally flat case appear in 
Bernstein-Gelfand-Gelfand (BGG) resolutions; see \cite{esl} and
references therein.  
Operators $D$ that exist in the conformally flat case and are not
{\em longest arrows} in even-dimensional 
BGG resolutions have {\em conformally
curved generalizations}; that is, operators $\tilde D$ that exist
and are covariant 
generally (without the assumption of conformal flatness), and which
generalize the conformally flat operator $D$.  The exceptional case,
longest arrows in even dimensions, consists 
exactly of operators
\beq\label{intertwinorwt}
\bbV^{r-n/2}(\l)\to\bbV^{-r-n/2}(\l)
\eeq
for which $(r,\tilde\l)$ is a strictly dominant integral or
half-integral $\gso(n+2)$ weight, and $n$ is even.  It is known
\cite{graham2} that the operator
$$
\bbV^1(0)\to\bbV^{-5}(0),
$$
which exists in the conformally flat case, has no conformally curved
generalization.  Since BGG resolutions are completely understood
in the conformally flat case, we know exactly when conformal covariants
exist in the conformally flat case. 

A {\em conformal covariant} is a natural differential operator
$D$ carrying sections of some $\bbV^s(\l)$ 
to sections of some $\bbV^t(\m)$ with 
$$
\bar g=\W^2g,\ 0<\W\in\ci\implies\bar D=\W^tD\W^{-s}.
$$
(The power of $\W$ on the far right is to be understood as a multiplication
operator.)  If the volume form $E$ and/or fundamental tensor-spinor $\g$ are 
involved, they are assumed to scale compatibly:
$$
\bar E=\W^nE,\ \bar\g=\W^{-1}\g.
$$
(The latter scaling is enforced by the Clifford relations.)

The discrete leading symbol may be viewed as a tool for converting
spectral information on differential operators into tensorial formulas.
In \cite{spg}, Sec.\ 3.a, 
a formula is given for the spectrum, on $S^n$, of intertwining
operators for the conformal group Spin$(n+1,1)$.  
A conformal covariant automatically gives rise to an intertwinor, and
the intertwinor with given weight parameters, if it exists, 
is unique (\cite{tbsw}, Sec.\ 6).  
Thus we have a formula for the spectrum of each conformal covariant
on each $\bbV(\l)$.
The problem, {\em a priori}, is that we have a formula for much more --
``most'' of the operators are only pseudo-differential, not differential, 
operators.  The spectral formula, from \cite{spg}, (3.3), is
$$
Z(r,\l)=\prod_{a=1}^{[(n+1)/2]}
\dfrac{\G(\ta_a+\frac12+r)}{\G(\ta_a+\frac12-r)}\,,
$$
as long as 
$$
n\ {\rm is\ odd\ or}\ \l_\ell\ne 0.
$$
The spectral function 
is to be viewed as a meromorphic function of $r$.  If a formula of this
type, as it stands, is identically zero or undefined, we renormalize by an
$\a$-independent meromorphic function of $r$ to bring out the information.
With this in mind, we could also consider
\beq\label{specfcn}
\tilde Z(r,\l)=\prod_{a\in\cT(\l)}
\dfrac{\G(\ta_a+\frac12+r)}{\G(\ta_a+\frac12-r)}\,,
\eeq
to be the spectral function, since the factors contributed by
$\cF(\l)$ just provide a meromorphic renormalization.  The order of
the resulting operator as a pseudo-differential operator is $2r$, and
the operator is intertwining between the spaces of
\nnn{intertwinorwt}.

In particular, to get a {\em differential} operator, it is necessary (but not
sufficient) that $2r$ be a nonnegative integer.  Given that $2r\in\bbN$, one
way to test whether our operator $D$ is differential is to try to 
write its spectrum
on $S^n$ as
a polynomial in the spectra of $G^*G$ (for gradients $G$), and, if applicable,
the self-gradient.  
If we happen to know (by BGG methods, say) that there is a differential
operator with these weight parameters, we can do something 
easier than realizing the spectrum as a polynomial in spectra of
low-order operators --
we can simply realize the discrete leading symbol as a polynomial
in discrete leading spectra of low-order operators.
If one of these processes succeeds, we have, in particular, a formula for the
principal part of $D$, and thus for $D$ on standard $\bbR^n$.  We can then 
use the formula for the conformal change of the Ricci tensor to write $D$
for any conformally flat metric.  In doing this, we can either (1) check BGG
resolutions to see whether there should be a covariant differential 
operator with the given parameters, or (2) simply go ahead with the procedure 
and see whether a covariant operator results.

More precisely, suppose we find that on $S^n$,
$$
D=P(\cD_u),
$$
where $\cD_u$ is $\dself$ when $G_u$ is a self-gradient, and is
$G_u^*G_u$
otherwise.  Now consider the conformal covariance relations for $G_u$, $G_u^*$,
and $\dself$: if $\bar g=e^{2\w}g$, then
$$
\begin{array}{rl}
\bar\dself&=\exp\left(-\frac{n+1}2\w\right)\dself 
\exp\left(\frac{n-1}2\w\right), \\
\bar G_u&=\exp\left(-\left(\frac{n+1}2+s_u\right)\w\right)G_u
\exp\left(\left(
\frac{n-1}2+s_u\right)\w\right), \\ 
\overline{G_u^*}&=\exp\left(-\left(\frac{n+1}2-s_u\right)\w\right)G_u^*
\exp\left(\left(\frac{n-1}2-s_u\right)\w\right).
\end{array} 
$$
Since
$$
\bar D=P(\bar\cD_u),
$$
this gives us a formula, involving $\w$, for $D$ at any conformally flat metric
in terms of the formula for $D$ at the standard flat metric.  

On the other hand, $\bar D$ is natural, 
so there should be a formula for it, in terms
of covariant derivatives in the metric $\bar g$, which does 
not explicitly mention
$\w$.  An efficient way to arrive at this formula is as follows.
Suppose $g_0$ is a flat metric, let $g_\w=e^{2\w}g_0$,
and affix the subscript $\w$ to all quantities computed in the metric $g_\w$.
Then
$$
\begin{array}{rl}
D_\w&=\exp\left(\left(-r-\frac{n}2\right)\w\right)D_0
\exp\left(\left(-r+\frac{n}2\right)\w\right) \\
&=\exp\left(\left(-r-\frac{n}2\right)\w\right)
P((G_u^*)_0(G_u)_0,(\dself)_0)\exp\left(\left(-r+\frac{n}2\right)\w\right) \\
&=
\exp\left(\left(-r-\frac{n}2\right)\w\right)
P\left(\exp\left(\left(\frac{n+1}2-s_u\right)\w\right)(G_u^*)_\w 
\exp\left(\left(2s_u+1\right)\w\right)
(G_u)_\w 
\exp\left(-\left(\frac{n-1}2+s_u\right)\w\right),\right. \\
&\left.{}\qquad\exp\left(\frac{n+1}2\w\right)
(\dself)_\w\exp\left(-\frac{n-1}2\w\right)\right)
\exp\left(\left(-r+\frac{n}2\right)\w\right).
\end{array}
$$
Now all covariant derivatives and curvatures 
involved in the expression on the far right are those of $g_\w$.  Applying the 
Leibniz rule to move $\w$ to the left, we 
obtain a natural differential operator
with coefficients that are polynomial in (in addition to the usual ingredients)
interated covariant derivatives of $\w$, of order at least 1.  
(The overall power of $e^\w$ in front is 0, since each $G^*G$
contributes $-2$, each $\dself$ contributes $-1$, and the homogeneity
degree of $P$ is $2r$.)

Now consider
the formula for the conformal change of the Ricci tensor, as applied to 
$g_0$ and $g_\w$.  
\begin{equation}\label{invtize}
\w_{ij}=-V_{ij}-\w_i\w_j+\frac12\w_k\w^k(g_\w)_{ij}\,.
\end{equation}
Here all covariant derivatives and curvatures are in the metric $g_\w$, and
$J$ and $V$ are the normalizations of the scalar curvature $K$ and
Ricci tensor $r$ that are best adapted to conformal geometry:
$$
J=\dfrac{K}{2(n-1)},\qquad V_{ij}=\dfrac{r_{ij}-Jg_{ij}}{n-2}\,.
$$
In \nnn{invtize}, we
have also employed the usual
notational abuse: for a scalar function, $\w_{j\ldots i}:=
\N_i\cdots\N_j\eta$.
Using \nnn{invtize} and its iterated covariant derivatives, we may
reduce the dependence of the coefficients to just $\N\w$.  The
condition that this dependence also disappears is equivalent to the
conformal covariance of $D$.

For example, consider the problem of finding a fourth-order conformal
covariant on trace-free symmetric 2-tensors.  We are assured of
the existence of such an operator $S$, in the conformally
flat case, by BGG considerations.  Substituting into the spectral
function, the discrete leading symbol of $S$ is
\beq\label{Sspec}
\dpsB(S)(q)=\left(q+\frac{n}2\right)\left(q+\frac{n}2-1\right)
\left(q+\frac{n}2-2\right)\left(q+\frac{n}2-3\right),\ \ 
q\in\{0,1,2\}.
\eeq
In this case, the parameter $\acirc$ has just one variable entry,
namely $\a_2$, which we have renamed $q$ for simplicity.

Note that by Corollary \ref{ellcor}, \nnn{Sspec} shows that $S$ is elliptic
whenever the dimension is not 2, 4, or 6.  It has positive definite
leading symbol when $n>6$, positive semidefinite leading symbol
when $n$ is 2, 4, or 6, and indefinite leading symbol when 
$n$ is 3 or 5.

By Sec.\ \ref{low}, we know that the discrete leading symbol of any natural
differential operator on $\TFS^2$ may be written as a polynomial
in the discrete leading symbols of $G_1^*G_1$ and $\N^*\N$, which
are the functions
$$
x(q):=\dfrac{(3-q)(q+n)}{3(n+2)}\ \ {\rm and}\ \ 1\ \ {\rm respectively}.
$$
We thus obtain a formula for the principal part of $S$ as a polynomial
in $G_1^*G_1$ and $\N^*\N$, that is,
$$
S=a(G_1^*G_1)^2+bG_1^*G_1\N^*\N+c(\N^*\N)+{\rm(lower\ order)},
$$
by simultaneously solving
$$
ax(q)^2+bx(q)+c=\left(q+\frac{n}2\right)\left(q+\frac{n}2-1\right)
\left(q+\frac{n}2-2\right)\left(q+\frac{n}2-3\right)
$$
for $q=0,1,2$.  The result is
$$
\bear{rl}
a&=9(n+2)^2, \\
b&=-\frac32(n+2)(n^2+6n+4), \\
c&=\frac1{16}n(n+2)(n+4)(n+6).
\eear
$$
In view of \nnn{preS}, we have an operator with principal part
\beq\label{Sppart}
\bear{l}
{}\!\!\!\!
B_{ij}\mapsto
\frac1{16}(n-2)n(n+2)(n+4)B_{ij|k}{}^k{}_l{}^l      
       -\frac12(n-2)n(n+2)B_{ik|j}{}^k{}_l{}^l             \\
       -\frac12(n-2)n(n+2)B_{jk|i}{}^k{}_l{}^l             
                   +2(n-2)nB_{kl|ij}{}^{kl}                 \\
                    +(n-2)nB_{kl|}{}^{kl}{}_m{}^mg_{ij}.     
\eear
\eeq
Applying the procedure described above to parlay the principal symbol
into a precise formula in the conformally flat case, we have:
\beq\label{long}
\begin{array}{l}
{}\!\!\!\!\!\!(\cS B)_{ij}=\frac1{16}(n-2)n(n+2)(n+4)B_{ij|k}{}^k{}_l{}^l      
       -\frac12(n-2)n(n+2)B_{ik|j}{}^k{}_l{}^l             \\
       -\frac12(n-2)n(n+2)B_{jk|i}{}^k{}_l{}^l             
                   +2(n-2)nB_{kl|ij}{}^{kl}                 \\
                    +(n-2)nB_{kl|}{}^{kl}{}_m{}^mg_{ij}     
-\frac1{16}(n-4)(n-2)n(n+2)(n+6)B_{ij|k}{}^kJ              \\
+\frac14{(n-2)n(n^2+2n-16)}B_{ik|j}{}^kJ                     
+\frac14{(n-2)n(n^2+2n-16)}B_{jk|i}{}^kJ                     \\
-\frac12{(n-2)n^2}B_{kl|}{}^{kl}g_{ij}J                     
+\frac1{64}(n-2)n(n+4)(n^3-2n^2-40n+64)B_{ij}J^2                     \\
+\frac18{(n-4)(n-2)n(n+2)}B_{jk|}{}^kJ_{|i}                 
+\frac18{(n-2)n(n^2-2n-40)}B_{ik|}{}^kJ_{|j}                     \\
-\frac1{16}{(n-2)n(n+2)(n^2-2n-56)}B_{ij|k}J_{|}{}^k        
-\frac18{(n-2)n(n^2+6n+40)}B_{ik|j}J_{|}{}^k                     \\
-\frac18{(n-2)n(n^2+6n+40)}B_{jk|i}J_{|}{}^k                
+\frac14{(n-4)(n-2)n(n+2)}B_{jk}J_{|i}{}^k                     \\
+\frac14{(n-6)(n-2)n(n+4)}B_{ik}J_{|j}{}^k                               
-\frac1{32}{(n-2)n(n+2)(n^2-48)}B_{ij}J_{|k}{}^k                     \\
-\frac12{(n-2)n^2}B_{kl}g_{ij}J_|{}^{kl}                    
-(n-4)(n-2)nB_{kl|}{}^{kl}V_{ij}                     \\
-2(n-6)(n-2)nB_{kl|}{}^kV_{ij|}{}^l                     
-n(n^2-10n+4)B_{kl}V_{ij|}{}^{kl}                     \\
+\frac14{(n-4)(n-2)n(n+2)}B_{jk|l}{}^lV_i{}^k            
-\frac18{(n-2)^2n(n^2-24)}B_{jk}JV_i{}^k                     \\
+\frac12{(n-4)(n-2)n(n+2)}B_{jk|l}V_i{}^k{}_|{}^l        
-(n-6)(n-2)nB_{kl|j}V_i{}^k{}_|{}^l                     \\
-2n(n+4)B_{kl}V_i{}^k{}_{|j}{}^l                      
+\frac14{(n-4)(n-2)n(n+2)}B_{jk|}{}^k{}_lV_i{}^l                     \\
-2(n-3)(n-2)nB_{kl|j}{}^kV_i{}^l                     
+\frac14{(n-4)(n-2)n(n+2)}B_{ik|l}{}^lV_j{}^k                     \\
-\frac18{(n-2)n(n^3-2n^2-24n+80)}B_{ik}JV_j{}^k       
+\frac12{(n-2)n(n^2-2n-16)}B_{ik|l}V_j{}^k{}_|{}^l                     \\
-(n-10)(n-2)nB_{kl|i}V_j{}^k{}_|{}^l                 
+2(n-2)nB_{kl}V_j{}^k{}_{|i}{}^l                     \\
+\frac14{(n-6)(n-2)n(n+4)}B_{ik|}{}^k{}_lV_j{}^l      
-2(n-5)(n-2)nB_{kl|i}{}^kV_j{}^l                     \\
+(n-2)n(n^2-8n+20)B_{kl}V_i{}^kV_j{}^l               
+\frac14{(n-2)n(n+2)(n+4)}B_{ij|kl}V^{kl}                     \\
-\frac14{(n-2)n(n^2+6n+16)}B_{ik|jl}V^{kl}            
-\frac14{(n-2)n(n^2+6n+16)}B_{jk|il}V^{kl}                     \\
+(n-2)n(n+8)B_{kl|ij}V^{kl}                     
-\frac12{(n-2)n^2}B_{kl|m}{}^mg_{ij}V^{kl}                     \\
+\frac14{(n-4)(n-2)n(n+4)}B_{kl}g_{ij}JV^{kl}    
-\frac12{(n-4)(n-2)n(n+8)}B_{kl}V_{ij}V^{kl}                     \\
+\frac18{(n-2)n(3n^3-2n^2-24n-32)}B_{jk}V_{il}V^{kl}  \\                   
+\frac18{(n-2)n(3n^3-2n^2-72n-128)}B_{ik}V_{jl}V^{kl}             \\
-\frac1{16}{(n-2)n(n+10)(n^2-4n-16)}B_{ij}V_{kl}V^{kl}        
-(n-2)^2nB_{kl|m}g_{ij}V^{kl}{}_|{}^m                     \\
+4(n-2)nB_{kl|}{}^k{}_mg_{ij}V^{lm}                     
-\frac14{(n-2)n(3n^2-6n-16)}B_{kl}g_{ij}V^k{}_mV^{lm}.                     
\end{array}
\eeq
(The transition from \nnn{Sppart} to \nnn{long} was accomplished
via an automated computation using Jack Lee's {\tt Ricci} package
\cite{ric}.)

Computing in the not necessarily conformally flat case (again using
{\tt Ricci}), one finds that the formula \nnn{long} is not conformally
covariant in general.  More precisely, the conformal variation
$$
\dfrac{d}{d\e}\Bigg|_{\e=0}
\left[\exp\left(\left(\frac{n}2+2\right)\e\eta\right)S_{\exp(2\e\eta)g}
\exp\left(-\left(\frac{n}2-2\right)\e\eta\right)\right]
$$
(where $\eta$ is an arbitrary smooth function) does not vanish
identically.  Schematically, this variation has three types of terms:
$(\N C)(\N\eta)B$, $C(\N\eta)\N B$, and $C(\N\N\eta)B$, where $C$ is
the Weyl conformal curvature tensor.  Using the conformally invariant
calculus of {\em tractors} \cite{beg}, however, Branson and Gover
\cite{tpbarg} have been able to get a formula for $S$ which is
conformally covariant 
in the general conformally curved case, and which reduces
to \nnn{long} for conformally flat metrics.

As another example, consider the bundle $\bbT$ of {\em twistors}, the
subbundle of the spinor-one-forms $\f_a$ with $\g^a\f_a=0$.  If $n$ is
odd, this is (isomorphic to) the irreducible bundle
$\bbV(\frac32,\frac12,\ldots,\frac12)$; if $n$ is even, $\bbT$ is
isomorphic to
\beq\label{twistsummands}
\bbV\left(\frac32,\frac12,\ldots,\frac12\right)
\os\bbV\left(\frac32,\frac12,\ldots,\frac12,-\frac12\right).
\eeq

Recall that the odd-dimensional case is worked out in detail above
(Example \ref{oddtwistor}):
the self-gradient $\cR$ 
is the {\em Rarita-Schwinger operator} \nnn{nmlrita}.
By BGG considerations, we are led to expect differential intertwinors
$\cR_{2r}$ carrying
$$
\bbT^{r-n/2}\to\bbT^{-r-n/2}
$$
for each positive, properly half-integral $r$.  That is, there is
an operator of each positive odd order.  By the spectral formula
\nnn{specfcn}, 
the spectrum of $\cR_{2r}$ on $S^n$ is (up to constant multiples)
$$
\bear{l}
\left\{(\ta_1-\frac12+r)(\ta_1-\frac32+r)\cdots(\ta_1+\frac12-r)
\right\}\cdot \\
\left\{(\ta_2-\frac12+r)(\ta_2-\frac32+r)\cdots(\ta_2+\frac12-r)
\right\}\cdot \\
\left\{(\ta_L-\frac12+r)(\ta_L-\frac32+r)\cdots(\ta_L+\frac12-r)
\right\},
\eear
$$
where $L=(n+1)/2$.
In particular, the discrete leading symbol is
$$
\bear{l}
\left\{(\ta_2-\frac12+r)(\ta_2-\frac32+r)\cdots(\ta_2+\frac12-r)
\right\}\cdot \\
\left\{(\ta_L-\frac12+r)(\ta_L-\frac32+r)\cdots(\ta_L+\frac12-r)
\right\},
\eear
$$
(In each $\cdots$, the factors decrease by 1 each time.)  The possible
values for $\ta_2$ are $\frac{n}2-1+q$, where $q\in\{0,1\}$, and
the possible values for $\ta_L$ are $\frac12\e$, where $\e=\pm 1$.  
Thus
the discrete leading symbol is
\beq\label{r3ds}
\bear{rl}
x(q,\e)&:=\left\{\left(\frac{n}2-\frac32+q+r\right)\cdots
\left(\frac{n}2-\frac12+q-r
\right)\right\}\cdot \\
&\left\{\left(\frac12\e-\frac12+r\right)\cdots\left(\frac12\e+\frac12-r
\right)\right\}
\eear 
\eeq
If we match $\e$-dependent factors symmetric about the middle one to
form differences of squares,
$$
\left(\frac12\e-\frac12+r-p\right)\left(\frac12\e+\frac12-r+p
\right)=\frac14\e^2-\left(\frac12-r+p\right)^2
=\frac14-\left(\frac12-r+p\right)^2,
$$
for $p$ running from 0 to $(2r-3)/2$, we get a polynomial in $r$
which does not depend on $\a$; thus, a renormalization in the 
sense described above.  The discrete leading symbol of the
renormalized operator is
$$
\bear{rl}
x(q,\e)&=\frac12\e
\left(\frac{n}2-\frac32+q+r\right)\cdots
\left(\frac{n}2-\frac12+q-r
\right) \\
&=\frac12\e(\frac{n}2-1+q)\prod_{p=1}^{(2r-1)/2}
\left\{(\frac{n}2-1+q+p)(\frac{n}2-1+q-p)\right\} \\
&=\frac12\e(\frac{n}2-1+q)\prod_{p=1}^{(2r-1)/2}
\left((\frac{n}2-1+q)^2-p^2\right).
\eear
$$

The self-gradient, being a conformal covariant, 
should be $\cR_1$, up to a constant factor, and thus, by
\nnn{r3ds},
should have discrete leading symbol $\frac12\e(\frac{n}2-1+q)$.
This agrees with \nnn{selfspec}.
If $\s$ is the discrete leading symbol of $\cR_1$, then \nnn{r3ds}
shows that the discrete leading symbol of $\cR_{2r}$ is
$$
\s\prod_{p=1}^{(2r-1)/2}(4\s^2-p^2),
$$
indicating that $\cR_{2r}$ has principal part
$$
\cR_1\prod_{p=1}^{(2r-1)/2}(4\cR_1^2-p^2\N^*\N).
$$
To see the precise normalization of $\cR_1$ in tensorial terms, 
note that by \nnn{selfspec}, \nnn{nmlrita}, and \nnn{cRita},
$$
\dfrac2{\sqrt{n(n+2)(n-2)}}(\cR_1\f)_i=
\left(\dfrac{n}{(n+2)(n-2)}\right)^{1/2}
\left\{\g^j\N_j\f_i-\frac2{n}\g_i\N^j\f_j\right\},
$$
so 
$$
(\cR_1\f)_i=
\dfrac{n}2
\left\{\g^j\N_j\f_i-\frac2{n}\g_i\N^j\f_j\right\}.
$$

Let us apply the procedure described above to get a formula for 
$\cR_3$ in the general conformally flat case.
We may re-express the principal part as a homogeneous polynomial
in 
$$
(\cR\f)_i=\g^j\N_j\f_i-(2/n)\g_i\N^j\f_j.
$$
and $\bbT\bbT^*$, where $\bbT$ is the 
{\em twistor operator} carrying spinors to twistors:
$$
(\bbT\y)_i=\N_i\y+(1/n)\g_i\g^j\N_j\y,
$$
since there is one linear relation among $\N^*\N$, $\cR^2$, and 
$\bbT\bbT^*$.
(The formal adjoint of $\bbT$ is $\bbT^*A=-\N^jA_j$.)  
The result is:
$$
\begin{array}{rl}
(\cR_3\f)_i&=\dfrac{n(n+2)}4(\cR^3\f)_i-\dfrac{4}{n-2}(\bbT\bbT^*\cR \f)_i \\
&\qquad
-\dfrac{n+2}{n}J\g_i\N^j\f_j+V_i{}^j\g^k\N_k\f_j+(n+2)V_i{}^k\g_k\N^j\f_j
+(n+1)V^{jk}\g_i\N_k\f_j \\
&\qquad-\dfrac{n(n+2)}{2}V^{jk}\g_k\N_j\f_i 
+(n-1)V^{jk}\g_k\N_i\f_j+V^{jl}\a_i{}^k{}_l\N_k\f_j
+\dfrac{n}{2}(\N^jJ)\g_i\f_j \\
&\qquad-\dfrac{n(n+2)}{4}(\N^jJ)\g_j\f_i
+n(\N^kV_i{}^j)\g_k\f_j,
\end{array}
$$
where 
$$
\a_{ijk}:=\g_{[i}\g_j\g_{k]}
$$
is the antisymmetrized iterated
Clifford symbol.
Again, the calculations were automated using {\tt Ricci}.

The operator and its conformal covariance relation may be considered
as polynomial identities in the dimension $n$.  Since the 
covariance relation holds for an infinite number of $n$ (all odd $n$),
it may be continued to even dimensions.  In the even dimensional case,
each $\cR_{2r}$ interchanges the summands in \nnn{twistsummands},
because $\cR_1$ does.

\end{document}